\documentclass[twocolumn,aps,superscriptaddress]{revtex4}

\usepackage{graphicx}
\usepackage{dcolumn}
\usepackage{bm}
\usepackage{amsmath,amssymb,amsfonts}
\usepackage{color}
\usepackage{braket}
\usepackage{mathtools}

\allowdisplaybreaks

\begin{document}

\title{Anomalous Hall effect from a non-Hermitian viewpoint}

\author{Hiroki Isobe}
\affiliation{RIKEN Center for Emergent Matter Science (CEMS), Wako, Saitama 351-0198, Japan}
\affiliation{Department of Applied Physics, University of Tokyo, Bunkyo, Tokyo 113-8656, Japan}

\author{Naoto Nagaosa}
\affiliation{RIKEN Center for Emergent Matter Science (CEMS), Wako, Saitama 351-0198, Japan}
\affiliation{Department of Applied Physics, University of Tokyo, Bunkyo, Tokyo 113-8656, Japan}

\begin{abstract}
Non-Hermitian descriptions often model open or driven systems away from the equilibrium.  Nonetheless, in equilibrium electronic systems, a non-Hermitian nature of an effective Hamiltonian manifests itself as unconventional observables such as a bulk Fermi arc and skin effects.  
We theoretically reveal that spin-dependent quasiparticle lifetimes, which signify the non-Hermiticity of an effective model in the equilibrium, induce the anomalous Hall effect, namely the Hall effect without an external magnetic field. 
We first examine the effect of nonmagnetic and magnetic impurities and obtain a non-Hermitian effective model.  
Then, we calculate the Kubo formula from the microscopic model to ascertain a non-Hermitian interpretation of the longitudinal and Hall conductivities.  
Our results elucidate the vital role of the non-Hermitian equilibrium nature in the quantum transport phenomena.  
\end{abstract}

\maketitle

\textit{Introduction}. 
A description of a material relies on a Hamiltonian.  For an electronic system, it describes the quantum-mechanical motion of electrons under a crystalline potential.  
The wavefunction in a clean system thus has a Bloch form, consisting of a plane wave and a short-range modulation by an underlying crystal.  
As a wave without decay, a Bloch function represents a current with the probability conserved, which is a consequence of the Hermiticity of the Hamiltonian.  
In reality, however, a Bloch wave is not an exact solution in the presence of impurities or disorder.  It decays during propagation, which we can effectively describe by a \textit{non-Hermitian} Hamiltonian \cite{Hatano1,Hatano2,Bender1,Bender2}.  
Examples of non-Hermitian effective models for quantum electronic systems include the electron-phonon coupling \cite{Kozii},  disorder \cite{Papaj,Zyuzin}, or strong correlation \cite{Yoshida,Kimura,Nagai}. 
In those systems, the non-Hermiticity causes a Fermi arc terminating with exceptional points or a drumhead-like flat band encircled by an exceptional ring \cite{Shen,Bergholtz}.  At an exceptional point, the non-Hermitian Hamiltonian is nondiagonalizable, which never appears from a Hermitian Hamiltonian \cite{Kato,Heiss}. 
Despite such observable spectral features, little has been known about the role of non-Hermiticity in a nonequilibrium, in particular a quantum transport phenomenon in solids \cite{Michishita,Michen,Michishita2}.

Non-Hermitian models appear in a variety of fields other than quantum systems \cite{Moiseyev,Rotter,Brody,Ashida}, such as photonics \cite{Guo,Lin,Feng,Regensburger,Peng,Wiersig,Chen,Hodaei,Hokmabadi,Zhen,Zhou1,Cerjan,Brandstetter,Xu,Zhao,Weidemann,photonics1,photonics2}, electrical circuits \cite{Hofmann1,Ezawa,Helbig,Hofmann2}, and mechanical systems \cite{Fleury,Makris1,Ding,Makris2,Rivet,Shi,Auregan,Brandenbourger,Scheibner,Zhou2,Rosa,Schomerus,Ghatak}.  In classical open or driven systems, non-Hermiticity arises from gain and loss, which accompanies the energy flow in and out of the system in focus.  It causes unusual features in spectrum, resonance, and propagation that never appear in a Hermitian model; e.g., sharp resonance and unidirectional transparency.  
Those resonance and wave propagation properties bring about advantages for measurements and detection through response of the system.  Therefore, they are easily observable as opposed to the spectrum features of quantum materials.

We investigate the linear response of a two-dimensional Dirac material with impurities.  We consider magnetic impurities in general, which induce \textit{spin-dependent} scattering, and derive an effective Hamiltonian from impurity averaging.  It reveals spin-dependent lifetimes leading to the non-Hermiticity.  
Independently, we evaluate the Kubo formula using the Dirac model with impurities by means of the conventional Feynman diagram technique.  
Our detailed calculations give the analytical expressions of the longitudinal and Hall conductivities, the latter of which emerges either from a uniform magnetization or randomly distributed spin-dependent impurities, along with the spin-orbit coupling embedded in the model.  
We reveal that the linear response properties manifest the non-Hermitian nature of the model; the spin-dependent lifetimes appearing in the effective Hamiltonian well approximate the longitudinal and Hall conductivities obtained from the Kubo formula.  
We also discuss the effect of skew scattering and the anomalous Hall effect induced by magnetic impurities without a uniform magnetization.

\textit{Model}. 
We consider the Dirac Hamiltonian in two dimensions
\begin{equation}
H_0(\bm{k}) = v\bm{k}\cdot\bm{\sigma} + m\sigma_z, 
\label{eq:Hamiltonian}
\end{equation}
where the Pauli matrices $\sigma_x$, $\sigma_y$, $\sigma_z$ represent the electron's spin, and $v$ and $m$ are the Dirac velocity and mass, respectively.  We set $\hbar = 1$ unless otherwise noted. 
We may regard the Hamiltonian as, e.g., a surface state of a topological insulator \cite{Qi,spin,Fu} with the mass $m$ corresponding to a uniform magnetization perpendicular to the plane induced by doping or deposition.  
For $m=0$, $H_0$ is invariant under time reversal $\mathcal{T} = i\sigma_y \mathcal{K}$ with the complex conjugation $\mathcal{K}$ as the mass $m$ renders a uniform background magnetization coupled to Dirac electrons via the exchange coupling.

We add impurities to the clean Dirac Hamiltonian.  While it is common to consider nonmagnetic potential impurities in massive models \cite{Onoda,Sinitsyn,Sinitsyn2,AHE,Ando1,Ando2}, impurities generally have potential and magnetic couplings concurrently \cite{Inoue,Kato2,Nunner1,Nunner2,Yang,Keser,Wakatsuki,EM}. 
In the following, we consider spin-dependent impurities, which break time-reversal symmetry, so that the clean Dirac Hamiltonian may be massless and time-reversal symmetric for a finite anomalous Hall effect.  
We assume here that each impurity has a magnetic moment perpendicular to the plane, which results in the impurity potential 
\begin{gather}
H_\text{imp}(\bm{r}) = V(\bm{r}) \eta, \quad
\eta = \eta_0 \sigma_0 + \eta_z \sigma_z =   
\begin{pmatrix}
\eta_{11} & 0 \\
0 & \eta_{22}
\end{pmatrix}.  
\end{gather}  
Finite $\eta_z$ describes the magnetic component of impurities.  It breaks time-reversal symmetry microscopically while the rotational symmetry in the $xy$ plane remains preserved.  The spin-dependent impurities have a correlation between the charge $(\sigma_0)$ and magnetic $(\sigma_z)$ sectors.  
For the impurity potential $V(\bm{r})$, we consider the moments of the spatial distribution 
\begin{equation}
\begin{aligned}
\langle V(\bm{r}) V(\bm{r}') \rangle &= \frac{n_i V_2}{(2\pi)^2} \delta(\bm{r}-\bm{r}') , \\
\langle V(\bm{r}) V(\bm{r}') V(\bm{r}'') \rangle &= \frac{n_i V_3}{(2\pi)^2} \delta(\bm{r}-\bm{r}') \delta(\bm{r}'-\bm{r}''), 
\end{aligned}
\end{equation}
where $\langle \cdot \rangle$ denotes impurity averaging and $n_i$ is the impurity concentration.  
$V_p$ $(p=2,3)$ represents a $p$-th order moment per single atomic potential.  
We set $\langle V(\bm{r}) \rangle = 0$ as the uniform component merely renormalizes the chemical potential and the mass.

\begin{figure}
\centering
\includegraphics[width=\hsize]{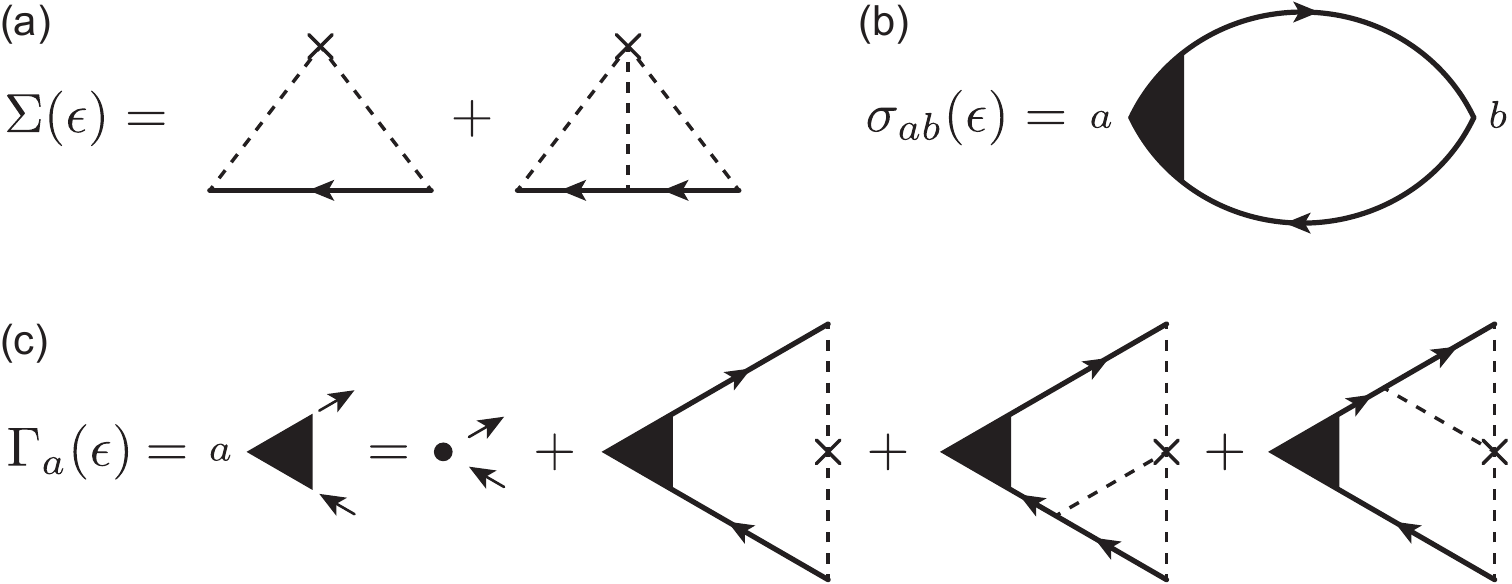}
\caption{Diagrammatic representations for the conductivity.  
(a) The self-energy captures the effect of time-reversal breaking in the effective non-Hermitian Hamiltonian.  A solid line represents the Green's function $G(\bm{k},\epsilon)$ and a cross and $p(=2,3)$ dashed lines correspond to the potential with the $p$-th moment. 
(b) The Kubo formula calculates the longitudinal and anomalous Hall conductivities.  
(c) The vertex correction describes the corrections to the scattering time by impurities.  
}
\label{fig:diagrams}
\end{figure}

\textit{Self-energy with impurity averaging}: 
We regard the impurity potential $H_\text{imp}(\bm{r})$ as a perturbation to the clean system $H_0(\bm{k})$.  We calculate the impurity average to obtain the self-energy $\Sigma^s(\epsilon)$, where $s=\text{R(A)}$ labels the retarded (advanced) function.  
The self-energy follows the self-consistent equation [Fig.~\ref{fig:diagrams}(a)], where the solid line represents the full Green's function $G^s(\bm{k},\epsilon) = [\epsilon - H_0(\bm{k}) - \Sigma^s(\epsilon)]^{-1}$ and a cross denotes an impurity. 
In the following, we focus on retarded functions as Hermitian conjugation gives the corresponding advanced functions. 
Without spontaneous symmetry breaking of the rotational symmetry, the retarded self-energy should have the form
\begin{equation}
\Sigma^\text{R}(\epsilon) = [\Sigma(\epsilon) - i\Gamma(\epsilon)] \sigma_0 + [\delta m(\epsilon) - i\gamma(\epsilon)] \sigma_z, 
\end{equation}
where $\Sigma(\epsilon)$, $\delta m(\epsilon)$, $\Gamma(\epsilon)$, and $\gamma(\epsilon)$ are real functions.  We henceforth refer to $\Sigma(\epsilon)$ and $\delta m(\epsilon)$ as real parts, and $\Gamma(\epsilon)$ and $\gamma(\epsilon)$ as imaginary parts.  
The real parts renormalize the energy and the mass as $\bar{\epsilon}(\epsilon) = \epsilon - \Sigma(\epsilon)$ and $\bar{m}(\epsilon) = m + \delta m(\epsilon)$, respectively.  
We obtain the explicit form of the self-energy later. 

\textit{Non-Hermitian effective Hamiltonian}. 
We define the retarded effective Hamiltonian after impurity averaging as 
\begin{equation}
H^\text{R}_\text{eff}(\bm{k},\epsilon) = H_0(\bm{k}) + \Sigma^\text{R}(\epsilon).  
\label{eq:effective}
\end{equation}
The effective Hamiltonian recovers translational symmetry, which is absent in the microscopic model due to the impurities $H_\text{imp}(\bm{r})$.  In compensation, the imaginary parts $\Gamma(\epsilon)$ and $\gamma(\epsilon)$ violate Hermiticity to describe the decay of Bloch waves. 
We note that Hermitian conjugation relates the retarded effective Hamiltonian not to itself but to the advanced one: $H^\text{A}_\text{eff}(\bm{k},\epsilon) = [H^\text{R}_\text{eff}(\bm{k},\epsilon)]^\dagger \neq H^\text{R}_\text{eff}(\bm{k},\epsilon)$.

The Dirac Hamiltonian is known to host the anomalous Hall effect with a finite mass, which requires time-reversal symmetry breaking.  It is beneficial to examine how the time-reversal operation $\mathcal{T} = i\sigma_y \mathcal{K}$ acts on the effective Hamiltonian.  Considering that retarded and advanced functions describe forward and backward time evolutions, a necessary condition for the microscopic model to have time-reversal symmetry is $\mathcal{T} H^\text{R}_\text{eff}(\bm{k},\epsilon) \mathcal{T}^{-1} = H^\text{A}_\text{eff}(-\bm{k},\epsilon)$, which we dub \textit{statistical} time-reversal symmetry for brevity; see Supplemental Material (SM) for details \cite{SM}.  
In the effective Hamiltonian \eqref{eq:effective}, $m(\epsilon)$ and $\gamma(\epsilon)$ break statistical time-reversal symmetry, where the former arises from a uniform magnetization and the latter from spin-dependent impurity scattering.  
We will see that both of them contribute to the anomalous Hall effect.  
We note that $\Gamma(\epsilon)$, which describes the spin-independent part of the quasiparticle lifetimes, does not break statistical time-reversal symmetry.  

Despite the non-Hermitian effective Hamiltonian, we can write the Green's function in the eigenstate basis.  As the effective Hamiltonian Eq.~\eqref{eq:effective} is non-Hermitian, we have distinct left and right eigenvectors $\bm{L}_s(\bm{k},\epsilon)$ and $\bm{R}_s(\bm{k},\epsilon)$ with $s=\pm$ corresponding to the two complex eigenvalues $E^\text{R}_s(\bm{k},\epsilon) = \Sigma(\epsilon) - i\Gamma(\epsilon) + s \sqrt{v^2 k^2 + [ m + \delta m(\epsilon) - i\gamma(\epsilon) ]^2}$ \cite{SM}.  The projection operator on an eigenstate is $P_s(\bm{k},\epsilon) = \bm{R}_s^T(\bm{k},\epsilon) \bm{L}_s(\bm{k},\epsilon)$. It is non-Hermitian and satisfies the completeness $\sum_{s=\pm} P_s(\bm{k},\epsilon) = \sigma_0$.  
Then, we obtain the Green's function \cite{SM}
\begin{equation}
\label{eq:projection}
G^\text{R}(\bm{k},\epsilon) = \frac{P_+(\bm{k},\epsilon)}{\epsilon-E^\text{R}_+(\bm{k},\epsilon)} + \frac{P_-(\bm{k},\epsilon)}{\epsilon-E^\text{R}_-(\bm{k},\epsilon)}.  
\end{equation}

\textit{Spin-dependent lifetimes}. 
Though a self-consistent solution requires a numerical calculation, we can analytically find a perturbative solution of the self-energy for weak impurities.  
As we will see later, the imaginary parts play an important role in the transport properties.  
The perturbative solutions of the imaginary parts are 
\begin{subequations}
\label{eq:gamma}
\begin{gather}
\begin{aligned}[b]
\Gamma(\epsilon) &\approx \frac{\alpha_2 \pi}{2} [ (\eta_{11}^2 + \eta_{22}^2) |\epsilon| + (\eta_{11}^2 - \eta_{22}^2) m \operatorname{sgn}(\epsilon) ] \\
&\quad - \frac{\pi \alpha_3 \Delta\Lambda}{\epsilon_0} [ (\eta_{11}^3 + \eta_{22}^3) |\epsilon| 
+ (\eta_{11}^3 - \eta_{22}^3) m \operatorname{sgn}(\epsilon) ], 
\end{aligned}\\
\begin{aligned}[b]
\gamma(\epsilon) &\approx \frac{\alpha_2 \pi}{2} [ (\eta_{11}^2 - \eta_{22}^2) |\epsilon| + (\eta_{11}^2 + \eta_{22}^2) m \operatorname{sgn}(\epsilon) ] \\
&\quad - \frac{\pi \alpha_3 \Delta\Lambda}{\epsilon_0} [ (\eta_{11}^3 - \eta_{22}^3) |\epsilon| 
+ (\eta_{11}^3 + \eta_{22}^3) m \operatorname{sgn}(\epsilon) ], 
\end{aligned}
\end{gather}
\end{subequations}
for $\epsilon^2 > m^2$.  
Here, we introduce the dimensionless constant for the impurity strength $\alpha_p = n_i^{p/2} V_p / (4\pi v)^{p/2}$ and the energy unit $\epsilon_0 = \sqrt{4\pi v^2 n_i}$.  
Since we have obtained the Green's function in the eigenstate basis Eq.~\eqref{eq:projection}, we impose energy cutoffs separately for the conduction and valence bands, $\Lambda_+$ and $\Lambda_-$, respectively.  
The difference of the energy cutoffs $\Delta\Lambda = \Lambda_+ - \Lambda_-$ appears at order $\alpha_3$.  
We retain the terms at order $\alpha_3$ because $\alpha_3 \Delta \Lambda / \epsilon_0$ can be comparable to $\alpha_2$ even for $|\alpha_3| \ll \alpha_2$ \cite{SM}.

Importantly, the present impurity model generates spin-dependent lifetimes, defined by the imaginary part of the self-energy as 
\begin{equation}
\tau_\uparrow = \frac{1}{2(\Gamma+\gamma)}, \quad 
\tau_\downarrow = \frac{1}{2(\Gamma-\gamma)}. 
\label{eq:lifetime}
\end{equation}
$\gamma(\epsilon)$ represents their difference, and importantly, it arises regardless of a uniform magnetization but by impurities with $\eta_{11} \neq \eta_{22}$, when the impurity scattering depends on spin.   
We note that the relation $\Gamma(\epsilon) \geq |\gamma(\epsilon)|$ must hold as the system is not driven by an external force.  In other words, the two lifetimes are positive and hence quasiparticles always decay.

\textit{Conductivity calculations}. 
Now we calculate the conductivity $\sigma_{ab}$ $(a,b=x,y)$ from the microscopic model $H_0 + H_\text{imp}$ using the Kubo formula.  
It is convenient to decompose the formula like as in the Kubo--St\v{r}eda formula \cite{Streda,Crepieux}, which leads to analytic solutions.  
At zero temperature \cite{temperature}, we write the electric conductivity at the Fermi level $\epsilon$ as $\sigma_{ab}(\epsilon) = \sigma^\text{(Ia)}_{ab}(\epsilon) + \sigma^\text{(Ib)}_{ab}(\epsilon) + \sigma^\text{(II)}_{ab}(\epsilon)$ [Fig.~\ref{fig:diagrams}(b)], 
where the three terms are 
$\sigma_{ab}^\text{(Ia)}(\epsilon) = \int_{\bm{k}} \operatorname{tr} [ j_a G^\text{R}(\epsilon) j_b G^\text{A}(\epsilon) ] /(2\pi)$, 
$\sigma_{ab}^\text{(Ib)}(\epsilon) = -\int_{\bm{k}} \operatorname{tr} [ j_a G^\text{R}(\epsilon) j_b G^\text{R}(\epsilon) 
+ j_a G^\text{A}(\epsilon) j_b G^\text{A}(\epsilon) ] /(4\pi)$, 
and 
$\sigma_{ab}^\text{(II)}(\epsilon) 
= \int_{\bm{k}} \int_{-\infty}^\epsilon d\epsilon' 
\operatorname{tr} [ j_a G^\text{R}(\epsilon') j_b \partial_{\epsilon'} G^\text{R}(\epsilon')
- j_a \partial_{\epsilon'} G^\text{R}(\epsilon') j_b G^\text{R}(\epsilon') 
+ j_a \partial_{\epsilon'} G^\text{A}(\epsilon') j_b G^\text{A}(\epsilon') 
- j_a G^\text{A}(\epsilon') j_b \partial_{\epsilon'} G^\text{A}(\epsilon') ] /(4\pi)$ with $\int_{\bm{k}} = \int d^2k/(2\pi)^2$. 
We omit $\bm{k}$ in the Green's function and the trace acts on the Pauli matrices for spin. 
$j_a$ is the current operator and its bare form without impurity scattering is $j_a = -ev\sigma_a$.  
Since we include the effect of scattering in the Green's function as a self-energy, we need to incorporate the vertex correction for a self-consistent calculation \cite{gauge_invariance}.  
In the following, we discuss the calculation of the conductivity at a low impurity concentration, i.e., consider an expansion with respect to $n_i$.  
Then, we should retain the vertex correction in $\sigma^\text{(Ia)}_{ab}$ while those in $\sigma^\text{(Ib)}_{ab}$ and $\sigma^\text{(II)}_{ab}$ give higher-order corrections \cite{Sinitsyn}.  We should thus replace one current operator $j_a$ in $\sigma^\text{(Ia)}_{ab}$ with $j_a = -ev \Gamma_a(\epsilon)$, which we should determine according to the self-consistent equation [Fig.~\ref{fig:diagrams}(c)] \cite{SM}.

By evaluating the Kubo formula, we obtain the analytic expression of the conductivity \cite{SM}
\begin{subequations}
\label{eq:exact}
\begin{gather}
\sigma^\text{(Ia)}_{ab}(\epsilon) = \frac{e^2}{4\pi^2} \left[ \bm{\Gamma}(\epsilon) \bm{\Lambda}(\epsilon) \right]_{ab}, \quad
\sigma^\text{(Ib)}_{ab}(\epsilon) = \frac{e^2}{4\pi^2} \delta_{ab}, \\
\sigma_{ab}^\text{(II)}(\epsilon) 
= -\frac{e^2}{4\pi^2} \varepsilon_{abz} \operatorname{Im} \ln \frac{\bar{\epsilon}-\bar{m}+i\Gamma+i\gamma}{\bar{\epsilon}+\bar{m}+i\Gamma-i\gamma}. 
\end{gather}
\end{subequations}
The matrices $\bm{\Gamma}(\epsilon)$ and $\bm{\Lambda}(\epsilon)$ are related to the vertex and ladder functions: 
\begin{gather}
\begin{aligned}[b]
\label{eq:vertex}
\bm{\Gamma} 
&= \{ \bm{1} - \alpha_2 \eta_{11}\eta_{22} \bm{\Lambda} \\
&\quad - \alpha_3 \eta_{11}\eta_{22} [ (\eta_{11} + \eta_{22}) \operatorname{Re}I_0^\text{R} + (\eta_{11} - \eta_{22}) \operatorname{Re}I_z^\text{R} ] \bm{\Lambda} \\
&\quad - \alpha_3 \eta_{11}\eta_{22} [ (\eta_{11} - \eta_{22}) \operatorname{Im}I_0^\text{R} + (\eta_{11} + \eta_{22}) \operatorname{Im}I_z^\text{R} ] \bm{\Lambda}\bm{\varepsilon}
\}^{-1}, 
\end{aligned}\\
\begin{aligned}[b]
\bm{\Lambda} 
&= \frac{\operatorname{Im}\ln\zeta}{\operatorname{Im}\zeta}
[ (\bar{\epsilon}^2+\Gamma^2-\bar{m}^2-\gamma^2) \bm{1} 
- 2(\bar{m}\Gamma+\bar{\epsilon}\gamma) \bm{\varepsilon} ], 
\end{aligned}
\end{gather}
where we use $(\bm{1})_{ab} = \delta_{ab}$ and $(\bm{\varepsilon})_{ab} = \varepsilon_{zab}$ with the Levi--Civita symbol $\varepsilon_{abc}$.  We also define the functions $\zeta(\epsilon) = (\bar{m}-i\gamma)^2 - (\bar{\epsilon}+i\Gamma)^2$ and 
\begin{subequations}
\label{eq:IR}
\begin{gather}
I_0^\text{R}(\epsilon) = - \frac{\Delta\Lambda}{\epsilon_0} - \frac{\bar{\epsilon}+i\Gamma}{\epsilon_0} \ln \frac{\Lambda_+ \Lambda_-}{\zeta}, \\
I_z^\text{R}(\epsilon) = -\frac{\bar{m}-i\gamma}{\epsilon_0} \ln \frac{\Lambda_+ \Lambda_-}{\zeta}.  
\end{gather}
\end{subequations}
We note that $\bm{\Gamma}(\epsilon)$, $\bm{\Lambda}(\epsilon)$, and $I^\text{R}_{0,z}(\epsilon)$ are dimensionless functions.

\begin{figure*}
\centering
\includegraphics[width=\hsize]{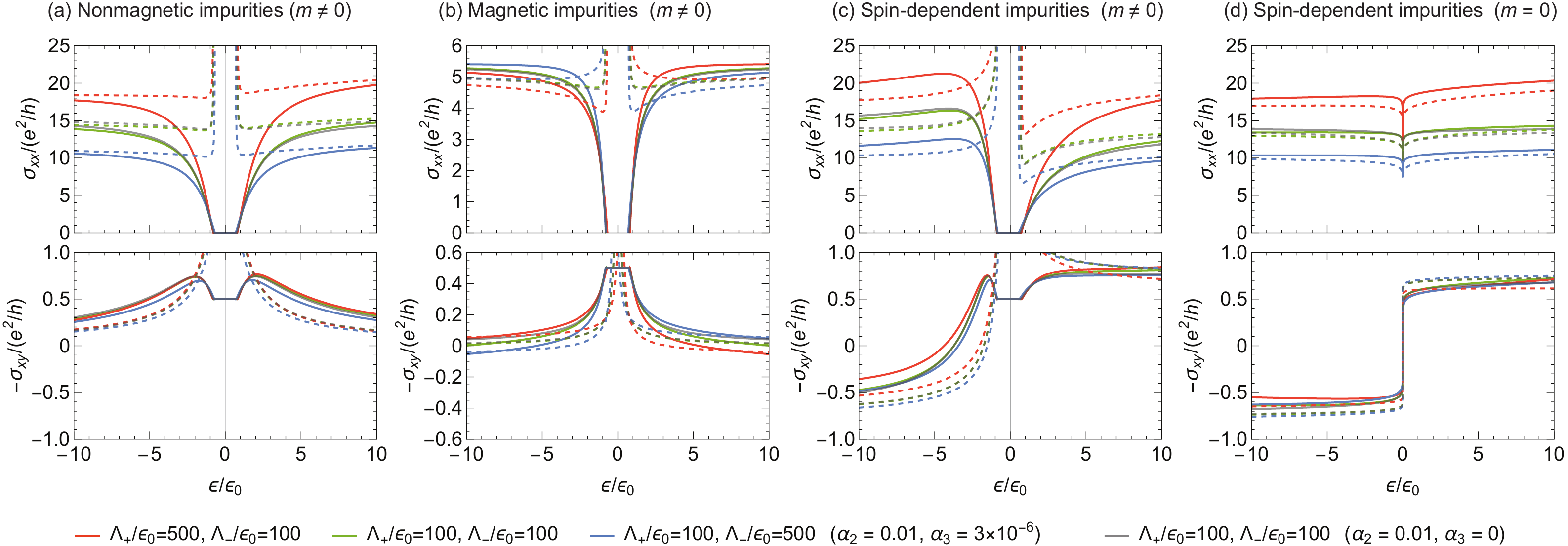}
\caption{
Fermi level dependence of the conductivity.  
The upper panels show the longitudinal conductivity $\sigma_{xx}$ and the lower panes the Hall conductivity $\sigma_{xy}$.  
(a)--(d) correspond to different impurity spin components and masses: (a) $\eta = \sigma_0$ (nonmagnetic), $m = \epsilon_0$ (massive); (b) $\eta = \sigma_z$ (magnetic), $m = \epsilon_0$ (massive); (c) $\eta_{11} = \sqrt{3/2}$, $\eta_{22} = \sqrt{1/2}$ (spin-dependent), $m = \epsilon_0$ (massive); and (d) $\eta_{11} = \sqrt{3/2}$, $\eta_{22} = \sqrt{1/2}$ (spin-dependent), $m = 0$ (massless).  We keep $\eta_{11}^2 + \eta_{22}^2$ constant for all cases.  
We choose the dimensionless constants for the impurity strength as $\alpha_2 = 0.01$, $\alpha_3 = 3 \times 10^{-6}$ for the colored lines, while the gray lines correspond to the cases without skew scattering $\alpha_2 = 0.01$, $\alpha_3 = 0$.  Different colors represent different energy cutoffs; see the legend.  The cutoff dependence is as weak as logarithmic for $\alpha_3 = 0$.  
The solid lines represent the exact solutions Eq.~\eqref{eq:exact} and the dashed lines the approximate results with a non-Hermitian interpretation Eq.~\eqref{eq:approximation}.  
}
\label{fig:conductivity}
\end{figure*}

$\sigma^\text{(Ib)}_{ab}$ contributes only to the longitudinal conductivity and $\sigma^\text{(II)}_{ab}$ to the Hall conductivity.  Roughly speaking, the Hall conductivity inside the band gap $(\bar{\epsilon}^2 < \bar{m}^2)$ comes from $\sigma^\text{(II)}_{ab}$ to give $\sigma_{xy} \approx -e^2/(2h)$ with the Planck constant $h(=2\pi\hbar)$ recovered. 
For large doping $|\epsilon| \gg |m|, \epsilon_0$, $\sigma^\text{(Ia)}_{ab}$ predominantly contributes to the conductivity.  For $|\alpha_3| \ll \alpha_2$ when skew scattering is not dominant, we find the approximate forms 
\begin{subequations}
\label{eq:approximation}
\begin{gather}
\sigma_{xx}(\epsilon) \approx \frac{e^2}{8\pi} \frac{|\epsilon|}{\Gamma(\epsilon)} \phi, \\
\sigma_{xy}(\epsilon) \approx -\frac{e^2}{4\pi} \frac{\gamma(\epsilon)}{\Gamma(\epsilon)} \phi^2 \operatorname{sgn}(\epsilon).  
\end{gather}
\end{subequations}
The constant $\phi = [ 1 - \eta_{11}\eta_{22} / (\eta_{11}^2+\eta_{22}^2) ]^{-1}$ originates from the vertex correction and hence characterizes transport quantities.  
From $\sigma_{xx}(\epsilon)$, we can identify $\tau_\text{tr}(\epsilon) = \phi/[2\Gamma(\epsilon)]$ as the transport scattering time \cite{Nomura,Guinea}.  
On the other hand, it is worth emphasizing that the approximate form of the anomalous Hall conductivity $\sigma_{xy}$ relies on $\gamma(\epsilon)$.  We recall that a finite $\gamma(\epsilon)$ manifests the spin-dependent lifetimes $(\tau_\uparrow \neq \tau_\downarrow)$ and the absence of time-reversal symmetry of the impurity model in the equilibrium.  
Using $\tau_{\uparrow,\downarrow}$, we find $\sigma_{xy} \propto (\tau_{\uparrow} - \tau_{\downarrow})/(\tau_{\uparrow} + \tau_{\downarrow})$ along with the spin-orbit coupling embedded in the Dirac model.  
The approximate forms Eq.~\eqref{eq:approximation} provide a non-Hermitian interpretation of the electric conductivity for both longitudinal and transverse components.

\textit{Numerical results}. 
We show the longitudinal conductivity $\sigma_{xx}$ and the Hall conductivity $\sigma_{xy}$ in Fig.~\ref{fig:conductivity}.  We evaluate the analytic expressions of the conductivity Eq.~\eqref{eq:exact} with the self-energy numerically obtained from the self-consistent equation [Fig.~\ref{fig:diagrams}(a)]. 
We use the conductivity unit $e^2/(2\pi\hbar) = e^2/h$ with the Planck constant $h$ recovered.
We present the results for various impurity types, masses, and energy cutoffs. 
We also depict the approximate results Eq.~\eqref{eq:approximation} in the same figure using the dashed lines, revealing a good agreement at relatively large doping from the Dirac point.  It corroborates the non-Hermitian interpretation of the longitudinal and Hall conductivities.  
The approximation deviates from the self-consistent solution for $|\epsilon| \lesssim |m|$, which coincides with the region where the semiclassical approximation breaks down.

The dimensionless parameter $\alpha_3$ characterizes the skewness of the impurity potential distribution whereas $\alpha_2$ does the impurity potential strength.  We note that $\alpha_3$ is a major source of skew scattering \cite{Smit1,Smit2}.  
With the symmetric energy cutoffs for the conduction and valence bands $(\Delta \Lambda = 0)$, the effect of skewness is tiny with the ratio $\alpha_3/\alpha_2^2 = 0.04$; compare the green and gray lines in Fig.~\ref{fig:conductivity}.  Its dependence is as weak as logarithmic, which we can infer from Eq.~\eqref{eq:IR}.
However, asymmetric cutoffs $(\Delta \Lambda \neq 0)$ enhance the effect of skewness (red and blue lines in Fig.~\ref{fig:conductivity}) as it appears with a potentially large factor $\Delta\Lambda/\epsilon_0$ in the self-energy Eq.~\eqref{eq:gamma} and the vertex correction Eq.~\eqref{eq:vertex}.  They modify the conductivity through the quasiparticle lifetime and the scattering time, respectively.

Now we discuss the effect of the magnetic properties of impurities.  For nonmagnetic impurities [Fig.~\ref{fig:conductivity}(a)], our result coincides with the previous result with symmetric energy cutoffs \cite{Sinitsyn,Ando1,Ando2,SM}.  In this case, a uniform magnetization that yields a finite mass breaks time-reversal symmetry to induce finite anomalous Hall effect.  The skewness $\alpha_3$ makes the conductivity asymmetric about the charge neutrality $\epsilon=0$ as it breaks electron-hole symmetry.  We observe a larger conductivity in the conduction band where $\alpha_3 \eta_0 \epsilon > 0$, because the scattering amplitude by an impurity is smaller when the impurity potential is repulsive \cite{Ando2,Novikov}.  
Also, we tend to observe a larger conductivity for $\Lambda_+ > \Lambda_-$, when the band in which the impurity potential is repulsive has a wider energy range.  
The peak structure of the conductivity implies broad resonance of scattering \cite{Ando2}, which is contained in the vertex correction Eq.~\eqref{eq:vertex} in the present analysis.  
For magnetic impurities [Fig.~\ref{fig:conductivity}(b)], the conductivity is reduced compared to the nonmagnetic impurity case with the same potential strength.  Here we observe certain electron-hole symmetry, which we will discuss later.

In reality, a magnetic impurity induces both potential and magnetic scatterings at a single site $(\eta_0, \eta_z \neq 0)$; in other words, impurity scattering becomes spin-dependent [Fig.~\ref{fig:conductivity}(c)].  
Then, the anomalous Hall effect appears even without a uniform magnetization $m = 0$ [Figs.~\ref{fig:conductivity}(d)].  The magnetism of impurities imparts time-reversal symmetry breaking, giving rise to $\delta m$ and $\gamma$; see Eq.~\eqref{eq:gamma} and SM \cite{SM}. 
The effect is prominent and realistic for $\eta_0, \eta_z \neq 0$ while purely magnetic impurities ($\eta_0=0$, $\eta_z \neq 0$) without magnetization $(m=0)$ can generate finite $\sigma_{xy}$ with $\alpha_3 \neq 0$ \cite{SM}.  The longitudinal conductivity shows a weak dependence on the Fermi level with a sharp dip at $\epsilon = 0$ to $\sigma_{xx} \simeq e^2/(2\pi^2)$ \cite{SM}.

\begin{figure}
\centering
\includegraphics[width=\hsize]{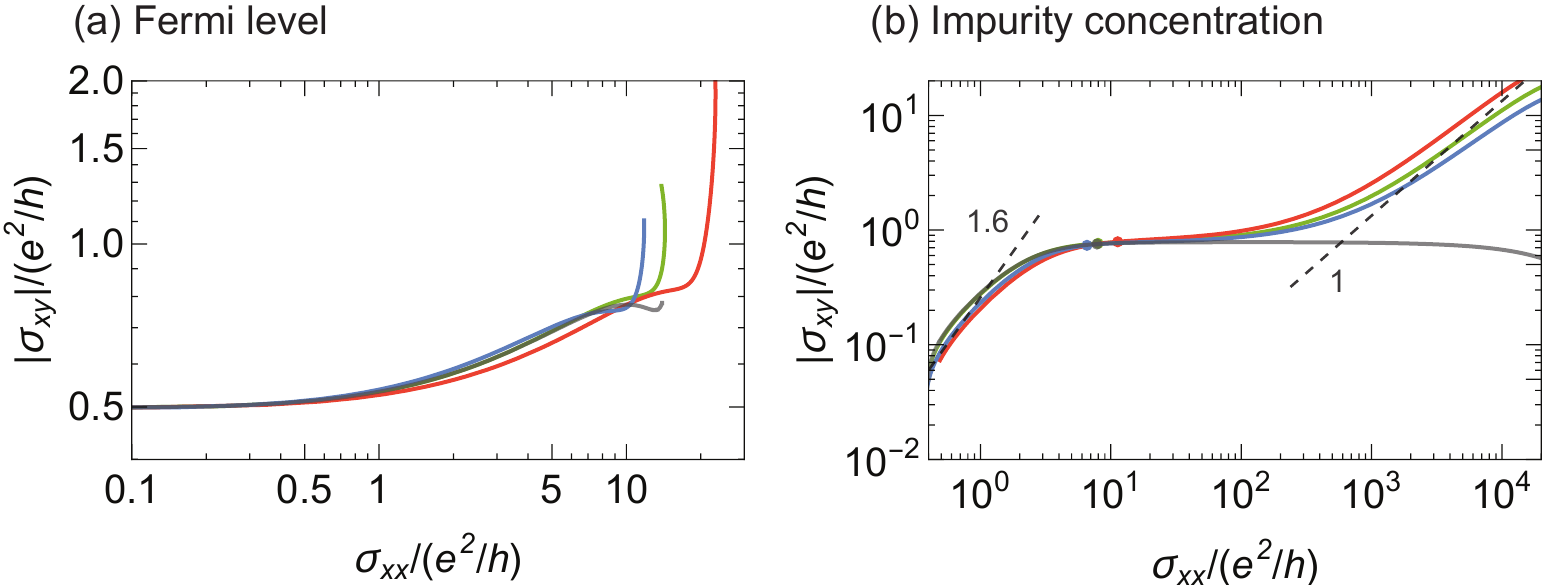}
\caption{Scaling relations between the longitudinal and Hall conductivities.  
We use the same color scheme as that of Fig.~\ref{fig:conductivity}.  
We show the scaling plots by varying (a) the Fermi level $(0 < \epsilon/\epsilon_0 < 100)$ and (b) the impurity concentration $n_i$ at $\epsilon/\epsilon_0 = 3$.  We choose spin-dependent impurities ($\eta_{11} = \sqrt{3/2}$, $\eta_{22}=\sqrt{1/2}$) and the mass $m=\epsilon_0$.  (a) corresponds to the positive energy region in Fig.~\ref{fig:conductivity}(c), and the points in (b) indicate the impurity concentration used in Fig.~\ref{fig:conductivity}(c).
}
\label{fig:scaling}
\end{figure}

\textit{Symmetries}. 
Some numerical results are symmetric or antisymmetric about the charge neutrality $(\epsilon = 0)$, which we can understand from the symmetries of the model.  We consider the following three symmetry operations: (i) time reversal $\mathcal{T} = i\sigma_y \mathcal{K}$, (ii) charge conjugation $\mathcal{C} = \sigma_x \mathcal{K}$, and (iii) their product $\mathcal{S} = \mathcal{T} \mathcal{C} = \sigma_z$.  
For convenience, we refer to $\mathcal{S}$ as ``sublattice'' symmetry \cite{Ryu}.  
$\mathcal{S}$ is a local operation acting on a spin, which we may view as reflection $(z \mapsto -z)$ about the two-dimensional system embedded in a three-dimensional space for the present model.  The clean Hamiltonian $H_0(\bm{k})$ has electron-hole symmetry
\begin{subequations}
\begin{equation}
\mathcal{C} H_0(\bm{k}) \mathcal{C}^{-1} = -H_0(-\bm{k}), 
\end{equation}
while the impurity potential transforms as 
\begin{equation}
\mathcal{C} V(\bm{r}) (\eta_0\sigma_0 + \eta_z\sigma_z) \mathcal{C}^{-1} = -V(\bm{r}) (-\eta_0\sigma_0 + \eta_z\sigma_z).  
\end{equation}
\end{subequations}
If $\eta_0 = 0$, the entire system preserves electron-hole symmetry $\mathcal{C} (H_0+H_\text{imp}) \mathcal{C}^{-1} = -(H_0+H_\text{imp})$.  
Therefore, the conductivity with magnetic impurities is symmetric about the charge neutrality as we have seen in Fig.~\ref{fig:conductivity}(b).  The operator $\mathcal{C}$ swaps the energy cutoffs for the conduction and valence bands as well.  
For $\eta_0 \neq 0$ and $\eta_z = 0$, the conductivity remains electron-hole symmetric if $\alpha_3 = 0$, since the distribution of the impurity potential retains electron-hole symmetry; see gray lines in Fig.~\ref{fig:conductivity}(a).  In other words, converting $V(\bm{r})$ to $-V(\bm{r})$ does not change $\alpha_2$; i.e., electron-hole symmetry is statistically preserved.

In the gapless case, on the other hand, $\mathcal{S}$ transforms the Hamiltonian as 
\begin{subequations}
\begin{gather}
\mathcal{S} H_0(\bm{k}) \mathcal{S}^{-1} = - H_0(\bm{k}) \quad (m=0), \\
\mathcal{S} H_\text{imp}(\bm{r}) \mathcal{S}^{-1} = H_\text{imp}(\bm{r}).  
\end{gather}
\end{subequations}
If $\alpha_3 = 0$, the model statistically preserves the ``electron-hole'' symmetry imposed by $\mathcal{S}$.  
As $\mathcal{S}$ is virtually a reflection about the plane, the Hall conductivity changes sign under $\mathcal{S}$ and hence it becomes antisymmetric about the charge neutrality whereas the longitudinal conductivity remains symmetric [Fig.~\ref{fig:conductivity}(d)].  
See SM for more details \cite{SM}.

\textit{Scaling}. 
Figure~\ref{fig:scaling} shows the scaling plots by varying the Fermi level $\epsilon$ and the impurity concentration $n_i$.  For other parameters, we use the same values as those for Fig.~\ref{fig:conductivity}(c).  
When we increase the Fermi level from the band edge [Fig.~\ref{fig:scaling}(a)], the longitudinal conductivity $\sigma_{xx}$ gradually increases while the Hall conductivity $\sigma_{xy}$ remains around $e^2/(2h)$.  For $e^2/h \lesssim \sigma_{xx} \lesssim 10 e^2/h$, there seems a scaling region with $|\sigma_{xy}| \propto \sigma_{xx}^{0.2}$ \cite{SM}.  As $\sigma_{xx}$ grows, it begins to saturate because of the artifact of short-range impurities \cite{Guinea}, but $\sigma_{xy}$ keeps growing linearly with the energy owing to skew scattering \cite{SM}, resulting in a rapid upturn in the scaling plot.  
On the other hand, the scaling plot by varying the impurity concentration reveals the known behavior.  As $\sigma_{xx}$ increases with smaller $n_i$, we observe the side-jump, intrinsic, and skew-scattering regions, where we find the approximate scaling relations $|\sigma_{xy}| \propto \sigma_{xx}^{1.6}$, $|\sigma_{xy}| \sim \text{const.}$, and $|\sigma_{xy}| \propto \sigma_{xx}^{1}$, respectively \cite{Onoda}.

\textit{Discussions}. 
In Fig.~\ref{fig:conductivity}(d), we observed the anomalous Hall effect in the absence of a uniform magnetization $(m=0)$.  It relies on the spin-dependent scattering $(\eta_0, \eta_z \neq 0)$, leading to spin-dependent lifetimes $\tau_\uparrow \neq \tau_\downarrow$ and thus non-Hermiticity of the effective model.  In reality, random magnetic impurities may have a finite uniform magnetization considering that the magnetic component $\eta_z$ arises from the exchange coupling.  We note that the gapped and gapless cases [Figs.~\ref{fig:conductivity}(c), (d)] are continuously connected.  
In addition, one might concern the violation of the Onsager reciprocal relation, when the Hall conductivity is finite without a uniform magnetization.  However, finite anomalous Hall effect requires magnetic impurities, which microscopically break time-reversal symmetry, so that our results comply with the Onsager reciprocal relation. 
Lastly, it is worth pointing out that the gapless Dirac model does not have a finite Berry curvature, so that it is natural to attribute finite $\sigma_{xy}$ to scattering-related phenomena rather than the intrinsic origin.  

\textit{Acknowledgments}. 
This work was supported by JST CREST Grant No. JPMJCR1874, Japan, and JSPS KAKENHI Grant No. 18H03676.

\onecolumngrid
\clearpage

\setcounter{section}{0}
\setcounter{equation}{0}
\setcounter{figure}{0}
\def\thesection{S\arabic{section}}
\def\theequation{S\arabic{equation}}
\def\thefigure{S\arabic{figure}}

\allowdisplaybreaks[3]

\begin{center}
{\bf\large
	Supplemental Material
}
\end{center}

In Supplemental Material (SM), we describe the non-Hermitian effective Hamiltonian, the calculation of the conductivity $\sigma_{ab}$, and the symmetry analysis of the model in detail.

\section{Model}

We consider the two-dimensional Dirac model
\begin{equation}
\label{eq:S_Hamiltonian}
H_0(\bm{k}) = v\bm{k}\cdot\bm{\sigma} + m\sigma_z, 
\end{equation}
where the Pauli matrices $\sigma_x$, $\sigma_y$, $\sigma_z$ designate the electron's spin.  
We set $\hbar = 1$ unless otherwise noted; we will recover the Planck constant $h$ when expressing the conductivity. 

We add impurities to the clean Hamiltonian to study the effect of scattering.  We suppose that each impurity has potential and magnetic components concurrently, resulting in the impurity Hamiltonian 
\begin{gather}
H_\text{imp}(\bm{r}) = V(\bm{r}) \eta.  
\label{eq:S_impurity}
\end{gather}
Here, we assume that the magnetic moments of the impurities point perpendicular to the plane, i.e., the spin matrix has the form 
\begin{equation}
\eta = \eta_0 \sigma_0 + \eta_z \sigma_z = 
\begin{pmatrix}
\eta_{11} & 0 \\
0 & \eta_{22}
\end{pmatrix}. 
\label{eq:S_eta}
\end{equation}
The impurities are randomly distributed in space.  We determine the properties of the impurity distribution by specifying the moments.  We assume no finite-range correlations and give the moments up to third order as 
\begin{equation}
\label{eq:S_moments}
\begin{aligned}
\langle V(\bm{r}) \rangle &= 0, \\
\langle V(\bm{r}) V(\bm{r}') \rangle &= \beta_2 \delta(\bm{r}-\bm{r}') , \\
\langle V(\bm{r}) V(\bm{r}') V(\bm{r}'') \rangle &= \beta_3 \delta(\bm{r}-\bm{r}') \delta(\bm{r}'-\bm{r}'').  
\end{aligned}
\end{equation}
The angle bracket $\langle \ \rangle$ stands for the average over the impurity distribution and the moments higher than fourth order are set zero.  We also set the first-order moment (mean) to be zero as it merely renormalizes the chemical potential and induces a uniform magnetization, which modifies the mass $m$.  
We write the constants $\beta_p$ using the impurity concentration $n_i$ as 
\begin{equation}
\label{eq:S_beta}
\beta_p = \frac{n_i V_p}{(2\pi)^2} \quad (p = 2,3).  
\end{equation}
We can regard the parameters $V_p$ as the $p$-th order moment per impurity.  We also define the dimensionless parameters as 
\begin{equation}
\label{eq:S_alpha}
\alpha_p = \frac{n_i^{p/2} V_p}{(4\pi v)^{p/2}} \quad (p = 2,3).  
\end{equation}

\section{Non-Hermitian effective Hamiltonian}

\subsection{Definition of the effective Hamiltonian}

The effects of impurities to the electrical conductivity are twofold when we evaluate it with the Kubo formula: one appears through the self-energy and the other is the vertex correction.  
We first calculate the self-energy, which arises as we take the impurity average.  Here, we treat the impurities perturbatively and consider the self-energy self-consistently.  

For a self-consistent treatment, we first need to postulate a resultant self-energy.  Now we use the short-ranged impurities with the spin component along the $z$ direction given by Eqs.~\eqref{eq:S_impurity} and \eqref{eq:S_eta}.  
Therefore, without spontaneous symmetry breaking of the rotational symmetry in the $xy$ plane, the self-energy should have the form 
\begin{gather}
\Sigma^\text{R}(\epsilon) = [\Sigma(\epsilon) - i\Gamma(\epsilon)] \sigma_0 + [\delta m(\epsilon) - i\gamma(\epsilon)] \sigma_z, \\
\Sigma^\text{A}(\epsilon) = [\Sigma(\epsilon) + i\Gamma(\epsilon)] \sigma_0 + [\delta m(\epsilon) + i\gamma(\epsilon)] \sigma_z.  
\end{gather}
The superscripts R and A represent the retarded and advanced functions, respectively.  
Given the self-energy, the full Green's function becomes 
\begin{gather}
G^\text{R}(\bm{k},\epsilon) = [ \epsilon \sigma_0 - H_0(\bm{k}) - \Sigma^\text{R}(\epsilon) ]^{-1}, \\
G^\text{A}(\bm{k},\epsilon) = [ \epsilon \sigma_0 - H_0(\bm{k}) - \Sigma^\text{A}(\epsilon) ]^{-1}.  
\end{gather}
The explicit form of the Green's function is 
\begin{gather}
G^\text{R}(\bm{k},\epsilon) = [ (\epsilon-\Sigma(\epsilon)+i\Gamma(\epsilon)) \sigma_0 - v\bm{k}\cdot\bm{\sigma} - (m+\delta m(\epsilon)-i\gamma(\epsilon)) \sigma_z ]^{-1}, \\
G^\text{A}(\bm{k},\epsilon) = [ (\epsilon-\Sigma(\epsilon)-i\Gamma(\epsilon)) \sigma_0 - v\bm{k}\cdot\bm{\sigma} - (m+\delta m(\epsilon)+i\gamma(\epsilon)) \sigma_z ]^{-1}.  
\end{gather}
The real parts of the self-energy, i.e., $\Sigma(\epsilon)$ and $\delta m(\epsilon)$, renormalize the energy and the mass, respectively.  We denote the renormalized energy and mass as 
\begin{gather}
\bar{\epsilon}(\epsilon) = \epsilon - \Sigma(\epsilon), \\
\bar{m}(\epsilon) = m + \delta m(\epsilon).  
\end{gather}
The Green's function thus becomes 
\begin{gather}
\label{eq:S_Green_R}
G^\text{R}(\bm{k},\epsilon) = [ (\bar{\epsilon}+i\Gamma) \sigma_0 - v\bm{k}\cdot\bm{\sigma} - (\bar{m}-i\gamma) \sigma_z ]^{-1}, \\
\label{eq:S_Green_A}
G^\text{A}(\bm{k},\epsilon) = [ (\bar{\epsilon}-i\Gamma) \sigma_0 - v\bm{k}\cdot\bm{\sigma} - (\bar{m}+i\gamma) \sigma_z ]^{-1}.  
\end{gather}

It is convenient to write the Green's function using the effective Hamiltonian 
\begin{gather}
G^\text{R}(\bm{k},\epsilon) = [ \epsilon\sigma_0 - H^\text{R}_\text{eff}(\bm{k},\epsilon) ]^{-1}, \\
G^\text{A}(\bm{k},\epsilon) = [ \epsilon\sigma_0 - H^\text{A}_\text{eff}(\bm{k},\epsilon) ]^{-1}, 
\end{gather}
where we define the effective Hamiltonian as 
\begin{gather}
\label{eq:S_Heff_R}
H^\text{R}_\text{eff}(\bm{k},\epsilon) = [ \Sigma(\epsilon) - i\Gamma(\epsilon) ] \sigma_0 + v\bm{k}\cdot\bm{\sigma} + [ m + \delta m(\epsilon) - i\gamma(\epsilon) ] \sigma_z, \\
\label{eq:S_Heff_A}
H^\text{A}_\text{eff}(\bm{k},\epsilon) = [ \Sigma(\epsilon) + i\Gamma(\epsilon) ] \sigma_0 + v\bm{k}\cdot\bm{\sigma} + [ m + \delta m(\epsilon) + i\gamma(\epsilon) ] \sigma_z.  
\end{gather}
Since the self-energy has the retarded and advanced components, we define the retarded and advanced effective Hamiltonians, respectively.  
The advanced effective Hamiltonian $H^\text{A}_\text{eff}(\bm{k},\epsilon)$ [Eq.~\eqref{eq:S_Heff_A}] is the Hermite conjugate of $H^\text{R}_\text{eff}(\bm{k},\epsilon)$ [Eq.~\eqref{eq:S_Heff_R}]: 
\begin{equation}
\label{eq:S_conjugate}
H^\text{A}_\text{eff}(\bm{k},\epsilon) = [H^\text{R}_\text{eff}(\bm{k},\epsilon)]^\dagger. 
\end{equation}  

We note that the definition of the effective Hamiltonian above does not receive an additional correction from the quasiparticle residue 
\begin{equation}
Z(\epsilon) = \left( 1 - \frac{\partial \Sigma(\epsilon)}{\partial\epsilon} \right)^{-1}.  
\end{equation}
Instead, the effective Hamiltonian defined above contains the energy dependence in addition to $\bm{k}$.  Physical observables such as the electrical conductivity of course do not depend on how we define the effective Hamiltonian; our definition is convenient for the following theoretical analyses and the interpretation of the outcomes.

\subsection{Eigenvalues and eigenvectors of the non-Hermitian effective Hamiltonian}

We here discuss the eigenvalues and eigenvectors of the retarded effective Hamiltonian $H^\text{R}_\text{eff}(\bm{k},\epsilon)$ [Eq.~\eqref{eq:S_Heff_R}].  
As the self-energy acquires finite imaginary components $\Gamma(\epsilon)$ and $\gamma(\epsilon)$, the effective Hamiltonian becomes non-Hermitian.  Unlike a Hermitian Hamiltonian, the Hermitian conjugate of the retarded effective Hamiltonian $H^\text{R}_\text{eff}(\bm{k},\epsilon)$ is not itself, but it becomes the advanced effective Hamiltonian $H^\text{A}_\text{eff}(\bm{k},\epsilon)$ as we have seen in Eq.~\eqref{eq:S_conjugate}.

A non-Hermitian Hamiltonian has different properties from those of a Hermitian counterpart.  One is that eigenvalues becomes complex in general and another is that we need to distinguish left and right eigenvalues.  
The retarded effective Hamiltonian \eqref{eq:S_Heff_R}, which is a $2 \times 2$ matrix, has the two complex eigenvalues 
\begin{equation}
E^\text{R}_\pm(\bm{k},\epsilon) = \Sigma(\epsilon) - i\Gamma(\epsilon) \pm \sqrt{v^2 k^2 + [ m + \delta m(\epsilon) - i\gamma(\epsilon) ]^2}.  
\end{equation}
Each eigenvalue has the corresponding left and right eigenvectors, $\bm{L}_\pm$ and $\bm{R}_\pm$, respectively, which satisfy 
\begin{gather}
H^\text{R}_\text{eff}(\bm{k},\epsilon) \bm{R}_\pm(\bm{k},\epsilon) = E^\text{R}_\pm (\bm{k},\epsilon) \bm{R}_\pm (\bm{k},\epsilon), \\
\bm{L}^T_\pm(\bm{k},\epsilon) H^\text{R}_\text{eff}(\bm{k},\epsilon) = \bm{L}^T_\pm(\bm{k},\epsilon) E^\text{R}_\pm(\bm{k},\epsilon).  
\end{gather}
We obtain the normalized left and right eigenvectors as 
\begin{gather}
\bm{L}_\pm(\bm{k},\epsilon) = \frac{1}{\sqrt{2}} \frac{1}{\sqrt{v^2 k^2 + (\bar{m}-i\gamma)^2 \pm (\bar{m}-i\gamma) \sqrt{v^2 k^2 + (\bar{m}-i\gamma)^2}}} 
\begin{pmatrix}
\bar{m}-i\gamma \pm \sqrt{v^2 k^2 + (\bar{m}-i\gamma)^2} \\ vk e^{-i\theta_{\bm{k}}}
\end{pmatrix}, \\
\bm{R}_\pm(\bm{k},\epsilon) = \frac{1}{\sqrt{2}} \frac{1}{\sqrt{v^2 k^2 + (\bar{m}-i\gamma)^2 \pm (\bar{m}-i\gamma) \sqrt{v^2 k^2 + (\bar{m}-i\gamma)^2}}} 
\begin{pmatrix}
\bar{m}-i\gamma \pm \sqrt{v^2 k^2 + (\bar{m}-i\gamma)^2} \\ vk e^{i\theta_{\bm{k}}}
\end{pmatrix}, 
\end{gather}
where $\theta_{\bm{k}}$ with the azimuthal angle of $\bm{k}$.
Here we define the normalization of the eigenvectors as 
\begin{gather}
\bm{L}_\pm (\bm{k},\epsilon) \cdot \bm{R}_\pm (\bm{k},\epsilon) = 1.  
\end{gather}

The two complex eigenvalues are distinct except when $\bar{m}(\epsilon) \neq 0$ and $v^2 k^2 \neq \gamma^2(\epsilon)$.  The degeneracy forms a ring in the two-dimensional $\bm{k}$ space, on which two eigenvectors $\bm{L}_\pm(\bm{k},\epsilon)$ or $\bm{R}_\pm(\bm{k},\epsilon)$ are no longer linearly independent.  This degeneracy collapses the projection operator that we use in the following analysis, which is known as an exceptional point specific to a non-Hermitian Hamiltonian.  Nevertheless, the volume of the singularity is negligible in the $\bm{k}$ space, so that it does not affect the calculations of the electrical conductivity.

In addition, the normalized eigenvectors satisfy the completeness 
\begin{equation}
\sum_{s = \pm} \bm{R}_s^T(\bm{k},\epsilon) \bm{L}_s(\bm{k},\epsilon) 
= \sum_{s = \pm} \bm{L}_s^T(\bm{k},\epsilon) \bm{R}_s(\bm{k},\epsilon) 
= \sigma_0, 
\label{eq:S_completeness}
\end{equation}
and the orthogonality
\begin{gather}
\bm{L}_s(\bm{k},\epsilon) \cdot \bm{R}_{s'}(\bm{k},\epsilon) = \delta_{ss'} \quad (s,s' = \pm), 
\end{gather}
away from the exceptional ring, where the two eigenvectors and eigenvalues coalesce.

\section{Self-energy}

\subsection{Self-consistent equations}

We now derive the self-consistent equations for the self-energy with the retarded Green's function \eqref{eq:S_Green_R} and the advanced Green's function \eqref{eq:S_Green_A}.  Since the impurity distribution has moment up to third order as we specified in Eq.~\eqref{eq:S_moments}, the self-consistent equation for the self-energy is 
\begin{align}
\label{eq:S_selfenergy}
\Sigma^r(\epsilon) &= 
n_i V_2 \eta \int_{\bm{k}} G^r(\bm{k},\epsilon) \eta 
+ n_i V_3 \eta \int_{\bm{k}} G^r(\bm{k},\epsilon) \eta \int_{\bm{k}'} G^r(\bm{k}',\epsilon) \eta \quad (r=\text{R,A}).  
\end{align}
We introduce the shorthand notation 
\begin{equation}
\int_{\bm{k}} = \int \frac{d^2k}{(2\pi)^2}. 
\end{equation}
To simplify the expression, we define the dimensionless quantity
\begin{equation}
\label{eq:S_functionI}
I^{r}(\epsilon) = \frac{4\pi v^2}{\epsilon_0} \int_{\bm{k}} G^{r}(\bm{k},\epsilon) 
= I_0^{r}(\epsilon) \sigma_0 + I_z^{r}(\epsilon) \sigma_z, 
\end{equation}
where $\epsilon_0$ is the energy unit defined by 
\begin{equation}
\epsilon_0 = \sqrt{4\pi v^2 n_i}.  
\end{equation}
We note that the function $I^{r}(\epsilon)$ contains the full Green's function, which depends on the self-energy $\Sigma^{r}(\epsilon)$.  

In the following, we only discuss the retarded component as we can obtain the corresponding advanced component by Hermite conjugation.  
The self-consistent equation for the retarded self-energy becomes 
\begin{align}
\Sigma^\text{R}(\epsilon) = \alpha_1 \epsilon_0 \eta + \alpha_2 \epsilon_0 \eta I^\text{R}(\epsilon) \eta +  \alpha_3 \epsilon_0 \eta I^\text{R}(\epsilon) \eta I^\text{R}(\epsilon) \eta.  
\end{align}
Using $I_0^\text{R}$ and $I_z^\text{R}$, we can transform the self-consistent equation into 
\begin{align}
\Sigma^\text{R}(\epsilon) &= 
\frac{1}{2} \alpha_1 \epsilon_0 [ (\eta_{11}+\eta_{22}) \sigma_0 + (\eta_{11}-\eta_{22}) \sigma_z ] \nonumber\\
&\quad + \frac{1}{2} \alpha_2 \epsilon_0 \{ [(\eta_{11}^2+\eta_{22}^2) I_0^\text{R}(\epsilon) + (\eta_{11}^2-\eta_{22}^2) I_z^\text{R}(\epsilon)] \sigma_0 + [ (\eta_{11}^2-\eta_{22}^2) I_0^\text{R}(\epsilon) + (\eta_{11}^2+\eta_{22}^2) I_z^\text{R}(\epsilon)] \sigma_z \} \nonumber\\
&\quad + \frac{1}{2} \alpha_3 \epsilon_0 \{ 
[ (\eta_{11}^3+\eta_{22}^3) [ (I_0^\text{R}(\epsilon))^2 + (I_z^\text{R}(\epsilon))^2 ] + 2 (\eta_{11}^3-\eta_{22}^3) I_0^\text{R}(\epsilon) I_z^\text{R}(\epsilon) ] \sigma_0 \nonumber\\
&\qquad\qquad\ + [ (\eta_{11}^3-\eta_{22}^3) [ (I_0^\text{R}(\epsilon))^2 + (I_z^\text{R}(\epsilon))^2 ] + 2 (\eta_{11}^3+\eta_{22}^3) I_0^\text{R}(\epsilon) I_z^\text{R}(\epsilon) ] \sigma_z 
\}.  
\label{eq:S_sc1}
\end{align}

The self-consistent equation is an integral equation to be solved at every $\epsilon$ since the function $I^\text{R}(\epsilon)$ includes the momentum integration.  The momentum integration diverges at large momenta, which we can easily see from a simple power counting.  
As the present model has the two energy bands, we can in general introduce two different cutoffs for the upper (conduction) band and the lower (valence) band.  We separate the Green's function into the two-band contributions using the band projection operator.  We define the projection operator based on the completeness Eq.~\eqref{eq:S_completeness} as 
\begin{align}
P_\pm(\bm{k},\epsilon) &= \bm{R}_\pm^T(\bm{k},\epsilon) \bm{L}_\pm(\bm{k},\epsilon) \nonumber\\
&= \frac{1}{2\sqrt{v^2 k^2 + (\bar{m}-i\gamma)^2}}
\begin{pmatrix}
\sqrt{v^2 k^2 + (\bar{m}-i\gamma)^2} \pm (\bar{m}-i\gamma) & \pm vke^{-i\theta_{\bm{k}}} \\
\pm vke^{i\theta_{\bm{k}}} & \sqrt{v^2 k^2 + (\bar{m}-i\gamma)^2} \mp (\bar{m}-i\gamma)
\end{pmatrix}.  
\end{align}
Using the projection operators, we obtain the Green's function as 
\begin{equation}
G^\text{R}(\bm{k},\epsilon) = \frac{P_+(\bm{k},\epsilon)}{\epsilon-E^\text{R}_+(\bm{k},\epsilon)} + \frac{P_-(\bm{k},\epsilon)}{\epsilon-E^\text{R}_-(\bm{k},\epsilon)}, 
\end{equation}
where the first and second terms on the right-hand side describe the conduction and valence band contributions, respectively.  
Now we impose the two cutoffs.  
While we originally have the momentum integral, the model has rotational symmetry and the result should depend only on the energy $\epsilon$.  Thus, we use the energy cutoffs $\Lambda_+$ and $\Lambda_-(>0)$ for the conduction and valence bands, respectively, instead of momentum cutoffs.  We note that the energy cutoffs are real-valued as the imaginary parts of the energy eigenvalues are comparably negligible at large energies.

We can perform the momentum integral in the function $I^\text{R}(\epsilon)$ to obtain 
\begin{gather}
I_0^\text{R}(\epsilon) = - \frac{\Lambda_+ - \Lambda_-}{\epsilon_0} - \frac{\bar{\epsilon}+i\Gamma}{\epsilon_0} L^\text{R}(\epsilon), \\
I_z^\text{R}(\epsilon) = -\frac{\bar{m}-i\gamma}{\epsilon_0} L^\text{R}(\epsilon), 
\end{gather}
where we introduce the functions 
\begin{gather}
\label{eq:S_functionL}
L^\text{R}(\epsilon) = 
\ln \frac{\Lambda_+ \Lambda_-}{\zeta(\epsilon)}, \\
\zeta(\epsilon) = (\bar{m}-i\gamma)^2 - (\bar{\epsilon}+i\Gamma)^2. 
\end{gather}
It depends logarithmically on the components of the self-energy $\Sigma^\text{R}(\epsilon)$.  We take the principal value of the complex logarithmic function.  
From the self-consistent equation \eqref{eq:S_sc1}, we obtain the coupled equations for the self-energy 
\begin{gather}
\begin{aligned}[b]
\Sigma(\epsilon) - i\Gamma(\epsilon) &= 
- \frac{1}{2} \alpha_2 (\eta_{11}^2+\eta_{22}^2) (\Lambda_+ - \Lambda_-) 
- \frac{1}{2} \alpha_2 [ (\eta_{11}^2+\eta_{22}^2) \bar{\epsilon} + (\eta_{11}^2-\eta_{22}^2) \bar{m} ] L^\text{R}(\epsilon) \\
&\quad + \frac{\alpha_3}{2\epsilon_0} \Big\{\! (\eta_{11}^3+\eta_{22}^3) \left[ (\Lambda_+-\Lambda_-)^2 + 2(\Lambda_+-\Lambda_-) \epsilon L^\text{R}(\epsilon) + (\bar{\epsilon}^2+\bar{m}^2) (L^\text{R}(\epsilon))^2 \right] \\
&\qquad\quad\quad + 2(\eta_{11}^3-\eta_{22}^3) \left[ (\Lambda_+-\Lambda_-) \bar{m} L^\text{R}(\epsilon) + \bar{\epsilon} \bar{m} (L^\text{R}(\epsilon))^2 \right] \!\Big\},
\end{aligned}\\
\begin{aligned}[b]
\delta m(\epsilon) - i\gamma(\epsilon) &= 
- \frac{1}{2} \alpha_2 [ (\eta_{11}^2-\eta_{22}^2) \bar{\epsilon} + (\eta_{11}^2+\eta_{22}^2) \bar{m} ] L^\text{R}(\epsilon) \\
&\quad + \frac{\alpha_3}{2\epsilon_0} \Big\{\! (\eta_{11}^3-\eta_{22}^3) \left[ (\Lambda_+-\Lambda_-)^2 + 2(\Lambda_+-\Lambda_-) \bar{\epsilon} L^\text{R}(\epsilon) + (\bar{\epsilon}^2+\bar{m}^2) (L^\text{R}(\epsilon))^2 \right] \\
&\qquad\quad\quad + 2(\eta_{11}^3+\eta_{22}^3) \left[ (\Lambda_+-\Lambda_-) \bar{m} L^\text{R}(\epsilon) + \bar{\epsilon} \bar{m} (L^\text{R}(\epsilon))^2 \right] \!\Big\}.  
\end{aligned}
\end{gather}
Some terms are independent of energy, which we can eliminate by redefining the energy $\epsilon$ and the mass $m$.  The self-consistent equation then becomes 
\begin{gather}
\label{eq:S_selfenergy_real}
\begin{aligned}[b]
\Sigma(\epsilon) - i\Gamma(\epsilon) &= 
- \frac{1}{2} \alpha_2 [ (\eta_{11}^2+\eta_{22}^2) \bar{\epsilon} + (\eta_{11}^2-\eta_{22}^2) \bar{m} ] L^\text{R}(\epsilon) \\
&\quad + \frac{\alpha_3}{2\epsilon_0} \Big\{\! (\eta_{11}^3+\eta_{22}^3) \left[ 2(\Lambda_+-\Lambda_-) \bar{\epsilon} L^\text{R}(\epsilon) + (\bar{\epsilon}^2+\bar{m}^2) (L^\text{R}(\epsilon))^2 \right] \\
&\qquad\quad\quad + 2(\eta_{11}^3-\eta_{22}^3) \left[ (\Lambda_+-\Lambda_-) \bar{m} L^\text{R}(\epsilon) + \bar{\epsilon} \bar{m} (L^\text{R}(\epsilon))^2 \right] \!\Big\},
\end{aligned}\\
\label{eq:S_selfenergy_imaginary}
\begin{aligned}[b]
\delta m(\epsilon) - i\gamma(\epsilon) &= 
- \frac{1}{2} \alpha_2 [ (\eta_{11}^2-\eta_{22}^2) \bar{\epsilon} + (\eta_{11}^2+\eta_{22}^2) \bar{m} ] L^\text{R}(\epsilon) \\
&\quad + \frac{\alpha_3}{2\epsilon_0} \Big\{\! (\eta_{11}^3-\eta_{22}^3) \left[ 2(\Lambda_+-\Lambda_-) \bar{\epsilon} L^\text{R}(\epsilon) + (\bar{\epsilon}^2+\bar{m}^2) (L^\text{R}(\epsilon))^2 \right] \\
&\qquad\quad\quad + 2(\eta_{11}^3+\eta_{22}^3) \left[ (\Lambda_+-\Lambda_-) \bar{m} L^\text{R}(\epsilon) + \bar{\epsilon} \bar{m} (L^\text{R}(\epsilon))^2 \right] \!\Big\}.  
\end{aligned}
\end{gather}
The difference of the energy cutoffs for the conduction and valence bands $\Lambda_+ - \Lambda_-$ appears at order $\alpha_3$.  Though $\alpha_3$ is much smaller than $\alpha_2$, $\alpha_3 (\Lambda_+ - \Lambda_-)$ might be comparable to $\alpha_2$, leading to an anomalously large contribution from the skewness of the impurity distribution.  

\begin{figure*}
\centering
\includegraphics[width=\hsize]{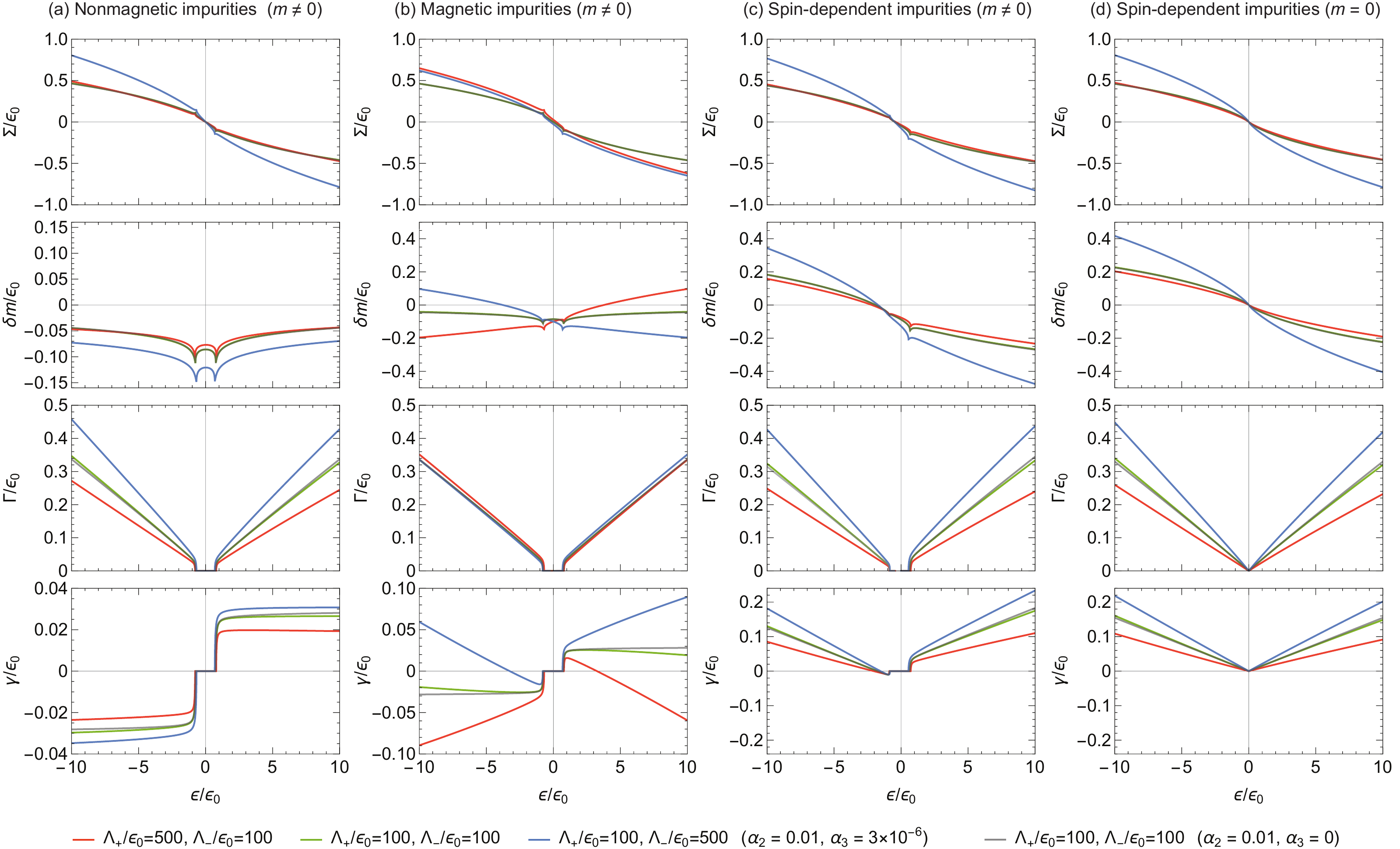}
\caption{
Energy dependence of the self-energy.  
The figure corresponds to the conductivity plot in Fig.~1 with the same parameter sets: (a) $\eta = \sigma_0$ (nonmagnetic), $m = \epsilon_0$ (massive); (b) $\eta = \sigma_z$ (magnetic), $m = \epsilon_0$ (massive); (c) $\eta_{11} = \sqrt{3/2}$, $\eta_{22} = \sqrt{1/2}$ (spin-dependent), $m = \epsilon_0$ (massive); and (d) $\eta_{11} = \sqrt{3/2}$, $\eta_{22} = \sqrt{1/2}$ (spin-dependent), $m = 0$ (massless).  
The dimensionless constants for the impurity strength are $\alpha_2 = 0.01$, $\alpha_3 = 3 \times 10^{-6}$ for the colored lines, and $\alpha_2 = 0.01$, $\alpha_3 = 0$ for the gray lines.  Different colors represent different energy cutoffs; see the legend.  
}
\label{fig:selfenergy}
\end{figure*}

In Fig.~\ref{fig:selfenergy}, we numerically solve the coupled self-consistent equations \eqref{eq:S_selfenergy_real} and \eqref{eq:S_selfenergy_imaginary} to obtain the self-energy.

\subsection{Perturbative solution at high energy}

The self-consistent equations for the self-energy require a numerical treatment in general.  In a specific situation, however, we can find an analytic solution.  
In this subsection, we consider the case where the energy $\epsilon$ is large enough, so that we can treat the self-energy induced by impurities perturbatively.  

We consider a perturbation theory with respect to the dimensionless parameters $\alpha_2$ and $\alpha_3$.  Here we use the unperturbed Green's function
\begin{equation}
G_0^\text{R}(\bm{k},\epsilon) = [ (\epsilon+i\delta)\sigma_0 - H_0(\bm{k}) ]^{-1},
\end{equation}
where $\delta$ is an infinitesimally small positive quantity.  
We use the unperturbed Green's function and evaluate the self-energy Eq.~\eqref{eq:S_selfenergy} to obtain the perturbative results.  
The unperturbed Green's function is equivalent to the limit where the self-energy is negligible: 
\begin{equation}
\label{eq:S_limit}
\Gamma \to +0, \quad 
\Sigma = \delta m = \gamma = 0. 
\end{equation}
We can thus utilize the previous calculations for the self-energy.  In the aforementioned limit, the function $L^\text{R}(\epsilon)$ becomes 
\begin{equation}
L^\text{R}(\epsilon) \to \ln \frac{\Lambda_+ \Lambda_-}{|m^2 - \epsilon^2|} + i\pi \operatorname{sgn}(\epsilon) \Theta(\epsilon^2 - m^2),  
\end{equation}
where $\Theta(x)$ is the unit step function.  
It does not contain the self-energy as we use the unperturbed Green's function instead of the full Green's function.  
As a result, we obtain the perturbative expression of the self-energy from Eqs.~\eqref{eq:S_selfenergy_real} and \eqref{eq:S_selfenergy_imaginary}: 
\begin{gather}
\label{eq:S_Sigma}
\begin{aligned}[b]
\Sigma(\epsilon) 
&\approx - \frac{1}{2} \alpha_2 \left[ (\eta_{11}^2+\eta_{22}^2) \epsilon + (\eta_{11}^2-\eta_{22}^2) m \right] \ln\frac{\Lambda_+ \Lambda_-}{|m^2 - \epsilon^2|} \\
&\quad + \frac{\alpha_3}{2\epsilon_0} \Bigg\{\! (\eta_{11}^3+\eta_{22}^3) \left[ 2(\Lambda_+-\Lambda_-) \epsilon \ln\frac{\Lambda_+ \Lambda_-}{|m^2 - \epsilon^2|} + (\epsilon^2+m^2) \left( \ln^2\frac{\Lambda_+ \Lambda_-}{|m^2 - \epsilon^2|} - \pi^2 \Theta(\epsilon^2-m^2) \right) \right] \\
&\qquad\quad\quad + 2(\eta_{11}^3-\eta_{22}^3) \left[ (\Lambda_+-\Lambda_-) m \ln\frac{\Lambda_+ \Lambda_-}{|m^2 - \epsilon^2|} + \epsilon m \left( \ln^2\frac{\Lambda_+ \Lambda_-}{|m^2 - \epsilon^2|} - \pi^2 \Theta(\epsilon^2-m^2) \right) \right] \!\bigg\}, 
\end{aligned}\\
\label{eq:S_delta_m}
\begin{aligned}[b]
\delta m(\epsilon) 
&\approx - \frac{1}{2} \alpha_2 \left[ (\eta_{11}^2-\eta_{22}^2) \epsilon + (\eta_{11}^2+\eta_{22}^2) m \right] \ln\frac{\Lambda_+ \Lambda_-}{|m^2 - \epsilon^2|} \\
&\quad + \frac{\alpha_3}{2\epsilon_0} \Bigg\{\! (\eta_{11}^3-\eta_{22}^3) \left[ 2(\Lambda_+-\Lambda_-) \epsilon \ln\frac{\Lambda_+ \Lambda_-}{|m^2 - \epsilon^2|} + (\epsilon^2+m^2) \left( \ln^2\frac{\Lambda_+ \Lambda_-}{|m^2 - \epsilon^2|} - \pi^2 \Theta(\epsilon^2-m^2) \right) \right] \\
&\qquad\quad\quad + 2(\eta_{11}^3+\eta_{22}^3) \left[ (\Lambda_+-\Lambda_-) m \ln\frac{\Lambda_+ \Lambda_-}{|m^2 - \epsilon^2|} + \epsilon m \left( \ln^2\frac{\Lambda_+ \Lambda_-}{|m^2 - \epsilon^2|} - \pi^2 \Theta(\epsilon^2-m^2) \right) \right] \!\bigg\}, 
\end{aligned}\\
\label{eq:S_Gamma}
\begin{aligned}[b]
\Gamma(\epsilon) &\approx \frac{1}{2} \alpha_2 \pi |\epsilon| \left[ (\eta_{11}^2+\eta_{22}^2) + (\eta_{11}^2-\eta_{22}^2) \frac{m}{\epsilon} \right] \Theta(\epsilon^2-m^2), \\
&\quad - \frac{\pi\alpha_3}{\epsilon_0} \Bigg\{\! (\eta_{11}^3+\eta_{22}^3) \left[ |\epsilon| (\Lambda_+-\Lambda_-) + (\epsilon^2+m^2) \ln\frac{\Lambda_+ \Lambda_-}{|m^2 - \epsilon^2|} \operatorname{sgn}(\epsilon) \right] \\
&\qquad\quad\quad + (\eta_{11}^3-\eta_{22}^3) m \left[ (\Lambda_+-\Lambda_-) + 2\epsilon \ln\frac{\Lambda_+ \Lambda_-}{|m^2 - \epsilon^2|} \right] \operatorname{sgn}(\epsilon) 
\!\Bigg\} \Theta(\epsilon^2-m^2), 
\end{aligned}\\
\label{eq:S_gamma}
\begin{aligned}[b]
\gamma(\epsilon) &\approx \frac{1}{2} \alpha_2 \pi |\epsilon| \left[ (\eta_{11}^2-\eta_{22}^2) + (\eta_{11}^2+\eta_{22}^2) \frac{m}{\epsilon} \right] \Theta(\epsilon^2-m^2) \\
&\quad - \frac{\pi\alpha_3}{\epsilon_0} \Bigg\{\! (\eta_{11}^3-\eta_{22}^3) \left[ |\epsilon| (\Lambda_+-\Lambda_-) + (\epsilon^2+m^2) \ln\frac{\Lambda_+ \Lambda_-}{|m^2 - \epsilon^2|} \operatorname{sgn}(\epsilon) \right] \\
&\qquad\quad\quad + (\eta_{11}^3+\eta_{22}^3) m \left[ (\Lambda_+-\Lambda_-) + 2\epsilon \ln\frac{\Lambda_+ \Lambda_-}{|m^2 - \epsilon^2|} \right] \operatorname{sgn}(\epsilon) 
\!\Bigg\} \Theta(\epsilon^2-m^2). 
\end{aligned}
\end{gather}

\subsection{Self-consistent solution at the charge neutrality in the massless model}
\label{sec:S_self-consistent-self-energy}

\subsubsection{Exact solution}

We can find the analytic solution of the self-consistent equation when electron-hole symmetry holds.  We define symmetry operators in Sec.~\ref{sec:S_symmetry}. 
To make the model electron-hole symmetric, the energy cutoffs for the conduction and valence bands must also be symmetric: $\Lambda_+ = \Lambda_- = \Lambda$.  Then, the present Dirac model satisfies electron-hole symmetry at $\epsilon = 0$.  Electron-hole symmetry forces the real part of the self-energy to vanish $\Sigma^\text{R}(0) = 0$.  We set $m = 0$ and $\alpha_3 = 0$ to further execute the analytic calculation for the imaginary part, where the self-consistent equation for the self-energy becomes 
\begin{align}
\Sigma^\text{R}(0) &= -i\Gamma\sigma_0 - i\gamma\sigma_z \nonumber\\
&= \frac{1}{2} \epsilon_0 \alpha_2 \ln\left(\frac{\Lambda^2}{\Gamma^2-\gamma^2}\right) \bigg[ \left( -(\eta_{11}^2+\eta_{22}^2) \frac{i\Gamma}{\epsilon_0} + (\eta_{11}^2-\eta_{22}^2) \frac{i\gamma}{\epsilon_0} \right) \sigma_0 + \left( -(\eta_{11}^2-\eta_{22}^2) \frac{i\Gamma}{\epsilon_0} + (\eta_{11}^2+\eta_{22}^2) \frac{i\gamma}{\epsilon_0} \right) \sigma_z \bigg].  
\end{align}
We can then solve this nonlinear equation analytically to obtain 
\begin{gather}
\Gamma = \Lambda \frac{|\eta_{11}|+|\eta_{22}|}{2\sqrt{|\eta_{11}\eta_{22}|}} \exp\left(-\frac{1}{2\alpha_2|\eta_{11}\eta_{22}|}\right), \\
\gamma = \Lambda \frac{|\eta_{11}|-|\eta_{22}|}{2\sqrt{|\eta_{11}\eta_{22}|}} \exp\left(-\frac{1}{2\alpha_2|\eta_{11}\eta_{22}|}\right). 
\end{gather}
We can find a similar calculation for a limited case in Ref.~\cite{S_Ostrovsky}.

Though the density of states vanishes at the charge neutrality without impurities, we find an exponentially small but finite imaginary part of the self-energy.  Particularly, finite $\gamma$ violate electron-hole symmetry, which indicates a spontaneous symmetry breaking at $\epsilon = 0$. 
The appearance of finite $\gamma$ induces exceptional points specific to a non-Hermitian Hamiltonian with $m=0$, which we discuss next.

\subsubsection{Effective Hamiltonian}

\begin{figure}
\centering
\includegraphics[width=0.4\hsize]{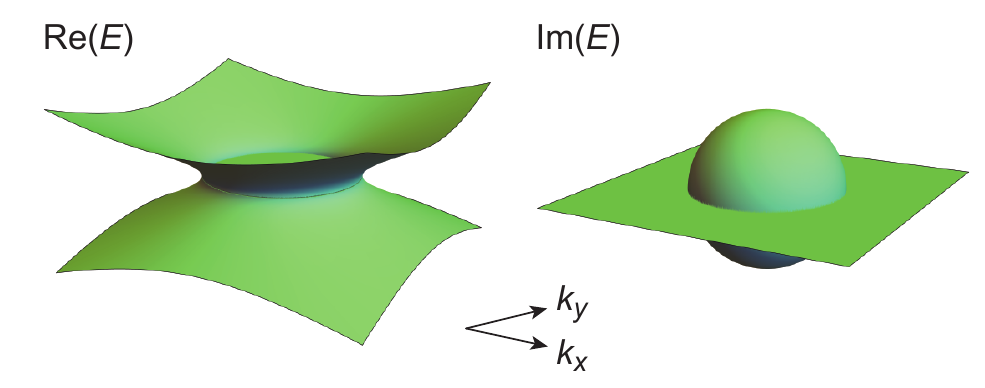}
\caption{Real and imaginary parts of the energy spectrum $E_\pm(\bm{k}) = -i\Gamma \pm \sqrt{v^2 k^2 - \gamma^2}$, corresponding to the effective Hamiltonian $H_\text{eff}^\text{R}(\bm{k}) = v\bm{k}\cdot\bm{\sigma} - i\Gamma \sigma_0 - i\gamma \sigma_z$. 
}
\label{fig:spectrum}
\end{figure}

In the vicinity of the charge neutrality, we may assume that $\Gamma$ and $\gamma$ are constant.  Then, the effective Hamiltonian is 
\begin{equation}
H_\text{eff}^\text{R}(\bm{k}) = v\bm{k}\cdot\bm{\sigma} + m\sigma_z - i\Gamma \sigma_0 - i\gamma \sigma_z, 
\end{equation}
and the two complex eigenvalues are 
\begin{equation}
E_\pm(\bm{k}) = -i\Gamma \pm \sqrt{v^2 k^2 + (m-i\gamma)^2}. 
\end{equation}
The real part of those energy eigenvalues has a drumhead shape at zero energy (Fig.~\ref{fig:spectrum}).  We find the degeneracy of the complex eigenvalues along $v^2 k^2 = \gamma^2$, and importantly the eigenvectors also coalesce along the ring, known as an exceptional line.  This singular behavior is specific to a non-Hermitian effective Hamiltonian.  
We emphasize, however, that the singular behavior does not affect our calculation of the conductivity, because we only utilize the non-Hermitian effective Hamiltonian to interpret the results of the conductivity.

\section{Vertex correction}

In this section, we calculate vertex corrections, namely corrections to the vertex operators or current operators reflecting impurity scattering in the present model.  While they have quantum-mechanical origins consisting of virtual intermediate states, there is a semiclassical interpretation related to transport scattering time.  
Since we consider later the current operators $j_a$ ($a = x,y$), we discuss the vertex operator $\Gamma^{rr'}_a(\epsilon)$ ($r,r' = \text{R},\text{A}$).  To maintain the consistency with the self-energy that we obtained earlier, the vertex function should read 
\begin{align}
\Gamma^{rr'}_a(\epsilon) &= \sigma_a 
+ n_i V_2 \eta \int_{\bm{k}} G^{r}(\bm{k},\epsilon) \Gamma^{rr'}_a(\epsilon) G^{r'}(\bm{k},\epsilon) \eta \nonumber\\
&\quad + n_i V_3 \eta \int_{\bm{k}\bm{k}'} G^{r}(\bm{k},\epsilon) \Gamma^{rr'}_a(\epsilon) G^{r'}(\bm{k},\epsilon) \eta G^{r'}(\bm{k}',\epsilon) \eta 
+ n_i V_3 \eta \int_{\bm{k}\bm{k}'} G^{r}(\bm{k}',\epsilon) \eta G^{r}(\bm{k},\epsilon) \Gamma^{rr'}_a(\epsilon) G^{r'}(\bm{k},\epsilon) \eta. 
\end{align}
We note that the Green's function $G^r(\bm{k},\epsilon)$ is the full Green's function including the self-energy.  The vertex function consists of $2 \times 2$ matrices, which we can decompose as 
\begin{equation}
\Gamma^{rr'}_a(\epsilon) = \Gamma^{rr'}_{ab}(\epsilon) \sigma_b.  
\end{equation}
Here, the summation over the repeated indices is implicit; the right-hand side of the equation indicates $\sum_{b = x,y} \Gamma^{rr'}_{ab}(\epsilon) \sigma_b$.  
The indices $a,b,c,d$ take $x,y$.  
One may wonder if the vertex function $\Gamma^{rr'}_a(\epsilon)$ could be expanded only with $\sigma_x$ and $\sigma_y$, which we shall confirm later.  

In proceeding the calculations, we introduce the quantity 
\begin{align}
\Lambda^{rr'}_a(\epsilon) &= 4\pi v^2 \int_{\bm{k}} G^{r}(\bm{k},\epsilon) \sigma_a G^{r'}(\bm{k},\epsilon) \nonumber\\
&= \Lambda^{rr'}_{ab}(\epsilon) \sigma_b,  
\end{align}
which corresponds to a ladder diagram.  
We also utilize the function $I^{r}(\epsilon)$, previously introduced in Eq.~\eqref{eq:S_functionI}.  Since retarded and advanced functions are related by Hermitian conjugation, we have the relations 
\begin{gather}
\int_{\bm{k}} G^\text{R}(\bm{k},\epsilon) = \frac{\epsilon_0}{4\pi v^2} I^\text{R}(\epsilon)
= \frac{\epsilon_0}{4\pi v^2} [I^\text{R}_0(\epsilon)\sigma_0 + I^\text{R}_z(\epsilon)\sigma_z], \\
\int_{\bm{k}} G^\text{A}(\bm{k},\epsilon) 
= \frac{\epsilon_0}{4\pi v^2} \{ [I^\text{R}_0(\epsilon)]^* \sigma_0 + [I^\text{R}_z(\epsilon)]^* \sigma_z \}. 
\end{gather}
We can now rewrite and simplify the vertex function using $\Lambda^{rr'}_{ab}(\epsilon)$ and $I^{r}(\epsilon)$.  We need to determine the two independent functions $\Gamma^\text{AR}_{ab}(\epsilon)$ and $\Gamma^\text{RR}_{ab}(\epsilon)$, while the others are related to the two by complex conjugation.  We find 
\begin{gather}
\label{eq:S_GammaAR}
\begin{aligned}[b]
\Gamma^{\text{AR}}_{ab}(\epsilon) \sigma_b &= \sigma_a + \alpha_2 \eta_{11}\eta_{22} \Gamma^{\text{AR}}_{ab}(\epsilon) \Lambda^{\text{AR}}_{bc}(\epsilon) \sigma_c \\
&\quad + \alpha_3 \eta_{11}\eta_{22} \Gamma^{\text{AR}}_{ab}(\epsilon) \Lambda^{\text{AR}}_{bc}(\epsilon) \{ [ (\eta_{11}+\eta_{22})\operatorname{Re}I^\text{R}_0(\epsilon) + (\eta_{11}-\eta_{22})\operatorname{Re}I^\text{R}_z(\epsilon) ] \sigma_c \\
&\hspace{114pt} + [ (\eta_{11}-\eta_{22})\operatorname{Im}I^\text{R}_0(\epsilon) + (\eta_{11}+\eta_{22})\operatorname{Im}I^\text{R}_z(\epsilon) ] (\varepsilon_{zcd} \sigma_d) \}, 
\end{aligned}\\
\label{eq:S_GammaRR}
\Gamma^\text{RR}_{ab}(\epsilon) \sigma_b = \sigma_a + \alpha_2 \eta_{11}\eta_{22} \Gamma^\text{RR}_{ab}(\epsilon) \Lambda^\text{RR}_{bc}(\epsilon) \sigma_c 
+ \alpha_3 \eta_{11}\eta_{22} \Gamma^\text{RR}_{ab}(\epsilon) \Lambda^\text{RR}_{bc}(\epsilon) [ (\eta_{11}+\eta_{22}) I^\text{R}_0(\epsilon) + (\eta_{11}-\eta_{22}) I^\text{R}_z(\epsilon) ] \sigma_c, 
\end{gather}
where $\varepsilon_{zab}$ is the Levi--Civita symbol.  

Now we evaluate the ladder function $\Lambda^{rr'}_{a}(\epsilon)$ to complete the calculation of the vertex function.  By performing momentum integrations, we obtain
\begin{gather}
\Lambda^\text{AR}_a(\epsilon) = \frac{\operatorname{Im}\ln\zeta(\epsilon)}{\operatorname{Im}\zeta(\epsilon)} 
[ (\bar{\epsilon}^2+\Gamma^2-\bar{m}^2-\gamma^2) \sigma_a - 2(\bar{m}\Gamma + \bar{\epsilon}\gamma) \varepsilon_{zab} \sigma_b ], \\
\Lambda^\text{RR}_a(\epsilon) = -\sigma_a.  
\end{gather}
We see that $\Lambda^\text{RR}_a(\epsilon)$ has a simple form, and we can easily obtain $\Gamma^\text{RR}_a(\epsilon)$ from Eq.~\eqref{eq:S_GammaRR} as 
\begin{align}
\Gamma^\text{RR}_a(\epsilon) &= \frac{1}{1 + \alpha_2 \eta_{11}\eta_{22} + \alpha_3 \eta_{11}\eta_{22} [ (\eta_{11}+\eta_{22}) I^\text{R}_0(\epsilon) + (\eta_{11}-\eta_{22}) I^\text{R}_z(\epsilon) ]} \sigma_a.  
\end{align}
$\Gamma^\text{RR}_a(\epsilon)$ acquires perturbatively small corrections to the bare value $\sigma_a$, which allows us to approximate \cite{S_Sinitsyn}
\begin{equation}
\Gamma^\text{RR}_a(\epsilon) \approx \sigma_a.  
\end{equation}
In contrast, $\Lambda^\text{AR}_a(\epsilon)$ increases linearly with respect to $\epsilon$ for large $|\epsilon|$ and its leading order term with respect to $\alpha_2$ is of order of $\alpha_2^{-1}$.    Therefore, the vertex correction has a significant effect.  In addition, $\Lambda^\text{AR}_{ab}(\epsilon)$ has off-diagonal elements, which are substantial for skew scattering.  

For a concise presentation of $2 \times 2$ matrices indexed by $a,b$, we introduce the matrix representation with boldface Greek symbols; here we define  
\begin{align}
[\bm{\Lambda}(\epsilon)]_{ab} &= \Lambda^\text{AR}_{ab}(\epsilon), \\
[\bm{\Gamma}(\epsilon)]_{ab} &= \Gamma^\text{AR}_{ab}(\epsilon).  
\end{align}
Then, $\bm{\Lambda}(\epsilon)$ and $\bm{\Gamma}(\epsilon)$ become 
\begin{gather}
\label{eq:S_functionLambda}
\bm{\Lambda}(\epsilon) = \frac{\operatorname{Im}\ln\zeta(\epsilon)}{\operatorname{Im}\zeta(\epsilon)} 
[ (\bar{\epsilon}^2+\Gamma^2-\bar{m}^2-\gamma^2) \bm{1} - 2(\bar{m}\Gamma + \bar{\epsilon}\gamma) \bm{\varepsilon} ], \\
\label{eq:S_functionGamma}
\begin{aligned}[b]
\bm{\Gamma}(\epsilon) 
&= \{ \bm{1} - \alpha_2 \eta_{11}\eta_{22} \bm{\Lambda}(\epsilon) \\
&\qquad - \alpha_3 \eta_{11}\eta_{22} [ (\eta_{11}+\eta_{22})\operatorname{Re}I_0^\text{R}(\epsilon) + (\eta_{11}-\eta_{22})\operatorname{Re}I_z^\text{R}(\epsilon) ] \bm{\Lambda}(\epsilon) \\
&\qquad - \alpha_3 \eta_{11}\eta_{22} [ (\eta_{11}-\eta_{22})\operatorname{Im}I_0^\text{R}(\epsilon) + (\eta_{11}+\eta_{22})\operatorname{Im}I_z^\text{R}(\epsilon) ] \bm{\Lambda}(\epsilon)\bm{\varepsilon} 
\}^{-1},  
\end{aligned}
\end{gather}
where we use the unit matrix $\bm{1}$ and a skew-symmetric matrix $\bm{\varepsilon}$, defined by 
\begin{equation}
\bm{1} = \begin{pmatrix} 1 & 0 \\ 0 & 1 \end{pmatrix}, \quad 
\bm{\varepsilon} = \begin{pmatrix} 0 & 1 \\ -1 & 0 \end{pmatrix}. 
\end{equation}
We note $\bm{\varepsilon} \bm{\varepsilon} = -\bm{1}$.  
Unlike $\Gamma^\text{RR}_a(\epsilon)$, we need to incorporate the off-diagonal elements and calculate the matrix inverse.  As we have mentioned, the term with $\alpha_2$ in $\bm{\Gamma}(\epsilon)$ is of order of unity, and hence we cannot treat it perturbatively.  On the other hand, the terms with $\alpha_3$ may be small depending on the choice of the energy cutoffs and the energy $\epsilon$.  However, they can be of order of unity when the energy cutoffs are highly asymmetric or the energy is close to the cutoff.  Therefore, we retain the expression as is and do not expand it with respect to $\alpha_3$ in evaluating $\bm{\Gamma}(\epsilon)$.

\section{Conductivity}

\subsection{Analytic expression of the electric conductivity}

With the self-energy and the vertex function obtained, we can easily calculate the electrical conductivity $\sigma_{ab}$ from the Kubo formula.  For an actual calculation, it is convenient to utilize the Kubo--St\v{r}eda formula, an expression to decompose the Kubo formula, to find an analytic solution.  The formula that we calculate is 
\begin{equation}
\sigma_{ab}(\epsilon) = \sigma^\text{(Ia)}_{ab}(\epsilon) + \sigma^\text{(Ib)}_{ab}(\epsilon) + \sigma^\text{(II)}_{ab}(\epsilon), 
\end{equation}
where the three constituent terms are 
\begin{gather}
\sigma_{ab}^\text{(Ia)}(\epsilon) = \frac{1}{2\pi} \int_{\bm{k}} \operatorname{tr} [ j_a G^\text{R}(\epsilon) j_b G^\text{A}(\epsilon) ], 
\\
\begin{aligned}[b]
\sigma_{ab}^\text{(Ib)}(\epsilon) = -\frac{1}{4\pi} \int_{\bm{k}} \operatorname{tr} [& j_a G^\text{R}(\epsilon) j_b G^\text{R}(\epsilon) 
+ j_a G^\text{A}(\epsilon) j_b G^\text{A}(\epsilon) ], 
\end{aligned}
\\
\begin{aligned}
\sigma_{ab}^\text{(II)}(\epsilon) 
&= \frac{1}{4\pi} \int_{\bm{k}} \int_{-\infty}^\epsilon d\epsilon' \operatorname{tr}\! \bigg[ j_a G^\text{R}(\epsilon') j_b \frac{dG^\text{R}(\epsilon')}{d\epsilon'} 
- j_a \frac{dG^\text{R}(\epsilon')}{d\epsilon'} j_b G^\text{R}(\epsilon') + j_a \frac{dG^\text{A}(\epsilon')}{d\epsilon'} j_b G^\text{A}(\epsilon') 
- j_a G^\text{A}(\epsilon') j_b \frac{dG^\text{A}(\epsilon')}{d\epsilon'} \bigg]. 
\end{aligned}
\end{gather}
Here, $j_a$ is the current operator, which we discuss in the next section, and we omit $\bm{k}$ as a variable of the Green's function for clarity.  
With the self-energy and the vertex corrections properly included, the expression above fulfills gauge invariance \cite{S_gauge_invariance}.  
The third term $\sigma^\text{(II)}_{ab}(\epsilon)$ contains the energy derivative of the Green's function.  We can utilize the Ward identity:  
\begin{equation}
\frac{dG^{r}(\bm{k},\epsilon)}{d\epsilon} = -[G^{r}(\bm{k},\epsilon)]^2, 
\end{equation}
leading to 
\begin{align}
\sigma_{ab}^\text{(II)}(\epsilon) 
&= \frac{1}{4\pi} \int_{\bm{k}} \int_{-\infty}^\epsilon d\epsilon' \operatorname{tr} [ - j_a G^\text{R}(\epsilon') j_b G^\text{R}(\epsilon') G^\text{R}(\epsilon')  
+ j_a G^\text{R}(\epsilon') G^\text{R}(\epsilon') j_b G^\text{R}(\epsilon') \nonumber\\
&\quad\hspace{72pt} - j_a G^\text{A}(\epsilon') G^\text{A}(\epsilon') j_b G^\text{A}(\epsilon') 
+ j_a G^\text{A}(\epsilon') j_b G^\text{A}(\epsilon') G^\text{A}(\epsilon') ]. 
\end{align}

We should determine the current operator.  
The bare expression of the current operator without an impurity correction is $-ev\sigma_a$, derived from the unperturbed Hamiltonian $H_0(\bm{k})$.  When we include the impurity correction, i.e., the vertex correction, the current operator is 
\begin{equation}
j_a = -ev \Gamma^{rr'}_a(\epsilon), 
\end{equation}
where we should choose $r,r'$ to match the Green's functions that appear before and after the current operator.  As we have discussed in the previous section, the vertex correction in $\Gamma^\text{RR}_a(\epsilon)$ and $\Gamma^\text{AA}_a(\epsilon)$ is negligible, but the corrections in $\Gamma^\text{AR}_a(\epsilon)$ and $\Gamma^\text{RA}_a(\epsilon)$ are significant.  In the following analysis, we retain the vertex correction only in $\Gamma^\text{AR}_a(\epsilon)$ or $\Gamma^\text{RA}_a(\epsilon)$.

We therefore obtain the expressions for the electric conductivity as 
\begin{gather}
\sigma_{ab}^\text{(Ia)}(\epsilon) = \frac{e^2 v^2}{2\pi} \int_{\bm{k}} \operatorname{tr} [ \Gamma^\text{AR}_a(\epsilon) G^\text{R}(\epsilon) \sigma_b G^\text{A}(\epsilon) ], 
\\
\sigma_{ab}^\text{(Ib)}(\epsilon) = -\frac{e^2 v^2}{4\pi} \int_{\bm{k}} \operatorname{tr} [ \sigma_a G^\text{R}(\epsilon) \sigma_b G^\text{R}(\epsilon) 
+ \sigma_a G^\text{A}(\epsilon) \sigma_b G^\text{A}(\epsilon) ], 
\\
\begin{aligned}[b]
\sigma_{ab}^\text{(II)}(\epsilon) 
&= \frac{e^2 v^2}{4\pi} \int_{\bm{k}} \int_{-\infty}^\epsilon d\epsilon' \operatorname{tr} [ -\, \sigma_a G^\text{R}(\epsilon') \sigma_b G^\text{R}(\epsilon') G^\text{R}(\epsilon')  
+ \sigma_a G^\text{R}(\epsilon') G^\text{R}(\epsilon') \sigma_b G^\text{R}(\epsilon') \\
&\quad\hspace{83pt} - \sigma_a G^\text{A}(\epsilon') G^\text{A}(\epsilon') \sigma_b G^\text{A}(\epsilon') 
+ \sigma_a G^\text{A}(\epsilon') \sigma_b G^\text{A}(\epsilon') G^\text{A}(\epsilon') ]. 
\end{aligned}
\end{gather}
Reflecting the preceding discussion, we neglect the vertex corrections in $\sigma^\text{(Ib)}_{ab}$ and $\sigma^\text{(II)}_{ab}$.

Using $\Lambda^\text{AR}(\epsilon)$ and $\Lambda^\text{RR}(\epsilon)$, we can evaluate $\sigma^\text{(Ia)}_{ab}(\epsilon)$ and $\sigma^\text{(Ib)}_{ab}(\epsilon)$ as follows: 
\begin{gather}
\label{eq:S_sigmaIa}
\begin{aligned}[b]
\sigma_{ab}^\text{(Ia)}(\epsilon) &= \frac{e^2 v^2}{2\pi} \Gamma^\text{AR}_{ac}(\epsilon) \int_{\bm{k}} \operatorname{tr} [ \sigma_c G^\text{R}(\epsilon) \sigma_b G^\text{A}(\epsilon) ] \\
&= \frac{e^2}{8\pi^2} \Gamma^\text{AR}_{ac}(\epsilon) \Lambda^\text{AR}_{cd}(\epsilon) \operatorname{tr} ( \sigma_d \sigma_b ) \\
&= \frac{e^2}{4\pi^2} \Gamma^\text{AR}_{ac}(\epsilon) \Lambda^\text{AR}_{cb}(\epsilon) \\
&= \frac{e^2}{4\pi^2} [\bm{\Gamma}(\epsilon)\bm{\Lambda}(\epsilon)]_{ab}, 
\end{aligned}\\
\label{eq:S_sigmaIb}
\begin{aligned}[b]
\sigma_{ab}^\text{(Ib)}(\epsilon) &= -\frac{e^2 v^2}{4\pi} \int_{\bm{k}} \operatorname{tr} [ \sigma_a G^\text{R}(\epsilon) \sigma_b G^\text{R}(\epsilon) 
+ j_\sigma G^\text{A}(\epsilon) \sigma_b G^\text{A}(\epsilon) ] \\
&= -\frac{e^2}{16\pi^2} \operatorname{tr} ( \Lambda^\text{RR}_a \sigma_b + \Lambda^\text{AA}_a \sigma_b ) \\
&= \frac{e^2}{4\pi^2} \delta_{ab}.
\end{aligned}
\end{gather}
We need some calculations for $\sigma^\text{(II)}_{ab}$, where we find 
\begin{align}
\sigma_{ab}^\text{(II)}(\epsilon) 
&= \frac{e^2 v^2}{4\pi} \int_{\bm{k}} \int_{-\infty}^\epsilon d\epsilon' \operatorname{tr} [ -\, \sigma_a G^\text{R}(\epsilon') \sigma_b G^\text{R}(\epsilon') G^\text{R}(\epsilon')  
+ \sigma_a G^\text{R}(\epsilon') G^\text{R}(\epsilon') \sigma_b G^\text{R}(\epsilon') \nonumber\\
&\quad\hspace{83pt} - \sigma_a G^\text{A}(\epsilon') G^\text{A}(\epsilon') \sigma_b G^\text{A}(\epsilon') 
+ \sigma_a G^\text{A}(\epsilon') \sigma_b G^\text{A}(\epsilon') G^\text{A}(\epsilon') ] \nonumber\\
&= \frac{e^2}{4\pi^2} \varepsilon_{abz} \int_{-\infty}^\epsilon d\epsilon' \left[ - \frac{i(\bar{m}-i\gamma)}{\zeta(\epsilon)} + \frac{i(\bar{m}+i\gamma)}{\zeta^*(\epsilon)} \right] \nonumber\\
&= \frac{e^2}{2\pi^2} \varepsilon_{abz} \int_{-\infty}^\epsilon d\epsilon' \operatorname{Im} \left[ \frac{\bar{m}-i\gamma}{\zeta(\epsilon')} \right] \nonumber\\
&= -\frac{e^2}{4\pi^2} \varepsilon_{abz} \operatorname{Im} \int_{-\infty}^\epsilon d\epsilon' \left( \frac{1}{\bar{\epsilon}-\bar{m}+i\Gamma+i\gamma} - \frac{1}{\bar{\epsilon}+\bar{m}+i\Gamma-i\gamma} \right) \nonumber\\
&\approx -\frac{e^2}{4\pi^2} \varepsilon_{abz} \operatorname{Im} \ln \frac{\bar{\epsilon}-\bar{m}+i\Gamma+i\gamma}{\bar{\epsilon}+\bar{m}+i\Gamma-i\gamma}. 
\label{eq:S_sigmaII}
\end{align}
On the last line, we neglect the energy dependence that appears through the self-energy to analytically execute the energy integration.  The resulting expression gives finite value mostly inside the energy gap $|\epsilon| \lesssim |m|$ where the self-energy is small, which justifies the approximation.

\subsection{Non-Hermitian interpretation}

We have obtained the analytic expression of the conductivity; however, one may feel somewhat complicated.  Then, one tends to insert the perturbative result of the self-energy and expand the expression with respect to the impurity strength and the impurity concentration.  It leads to an explicit result as a function of the energy, whereas its implication remains vague to some extent.  

Here we aim to simplify the result while retaining the self-energy as is.  To this end, we first examine the approximate expressions of the ladder and vertex functions Eqs.~\eqref{eq:S_functionLambda} and \eqref{eq:S_functionGamma}.  At a large energy $|\epsilon| \gg |m|, |\Sigma(\epsilon)|, |\delta m(\epsilon)|, \Gamma(\epsilon), |\gamma(\epsilon)|$, and for weak skew scattering $\alpha_2 \gg |\alpha_3 \epsilon/\epsilon_0|, |\alpha_3 (\Lambda_+-\Lambda_-)/\epsilon_0|$, we obtain the ladder and vertex functions 
\begin{gather}
\bm{\Lambda}(\epsilon) \approx \frac{\pi \operatorname{sgn}(\epsilon)}{2\epsilon\Gamma(\epsilon)} 
[ \epsilon^2 \bm{1} - 2\epsilon \gamma(\epsilon) \bm{\varepsilon} ], \\
\begin{aligned}[b]
\bm{\Gamma}(\epsilon) 
&\approx [ \bm{1} - \alpha_2 \eta_{11}\eta_{22} \bm{\Lambda}(\epsilon) ]^{-1} \\
&\approx \left[ \left( 1 - \frac{\pi}{2} \alpha_2 \eta_{11} \eta_{22} \frac{|\epsilon|}{\Gamma(\epsilon)} \right) \bm{1} + \pi \alpha_2 \eta_{11} \eta_{22} \frac{\gamma(\epsilon)}{\Gamma(\epsilon)} \operatorname{sgn}(\epsilon) \bm{\varepsilon} \right]^{-1}, 
\end{aligned}
\end{gather}
where we can approximate $\zeta(\epsilon) \approx -\epsilon^2 - 2i \epsilon \Gamma(\epsilon)$ and utilize the fact that $\Gamma(\epsilon)$ and $\gamma(\epsilon)$ are of order of $\alpha_2 \epsilon$.  
We define the function 
\begin{equation}
\tilde{\phi}(\epsilon) = \left( 1 - \frac{\pi}{2} \alpha_2 \eta_{11} \eta_{22} \frac{|\epsilon|}{\Gamma(\epsilon)} \right)^{-1} 
\end{equation}
to clarify the contribution from the vertex correction.  Since we consider the high doping regime, $\sigma^\text{(Ia)}_{ab}$ is predominant and hence we obtain
\begin{align}
\sigma_{ab}(\epsilon) &\approx \frac{e^2}{4\pi^2} \left[ \bm{\Gamma}(\epsilon) \bm{\Lambda}(\epsilon) \right]_{ab} \nonumber\\
&\approx \left[ \frac{e^2}{8\pi} \frac{|\epsilon|}{\Gamma(\epsilon)} \tilde{\phi}(\epsilon) \bm{1} - \frac{e^2}{4\pi} \frac{\gamma(\epsilon)}{\Gamma(\epsilon)} \operatorname{sgn}(\epsilon) \tilde{\phi}^2(\epsilon) \bm{\varepsilon} \right]_{ab}.  
\end{align}

We can simplify the function $\tilde{\phi}(\epsilon)$ under the current assumptions, which leads with Eq.~\eqref{eq:S_Gamma} to 
\begin{equation}
\tilde{\phi}(\epsilon) \approx \left( 1 - \frac{\eta_{11}\eta_{22}}{\eta_{11}^2 + \eta_{22}^2} \right)^{-1} \equiv \phi.  
\end{equation}
We can see that it is approximately constant independent of the energy $\epsilon$.  
Therefore, the approximate forms of the longitudinal conductivity and the Hall conductivity become 
\begin{gather}
\label{eq:S_sigmaxx_NH}
\sigma_{xx}^\text{NH}(\epsilon) = \frac{e^2}{8\pi} \frac{|\epsilon|}{\Gamma(\epsilon)} \phi, \\
\label{eq:S_sigmaxy_NH}
\sigma_{xy}^\text{NH}(\epsilon) = - \frac{e^2}{4\pi} \frac{\gamma(\epsilon)}{\Gamma(\epsilon)} \phi^2 \operatorname{sgn}(\epsilon), 
\end{gather}
respectively.  
The last expressions have a clear interpretation from a non-Hermitian viewpoint.  $\Gamma(\epsilon)$ and $\gamma(\epsilon)$ represent the spin-independent and spin-dependent parts of the quasiparticle lifetime, respectively, making the effective Hamiltonian non-Hermitian.  As they are equilibrium quantities, we need the vertex correction represented by the factor $\phi$ for the transport quantities.

\subsection{Numerical results}

\begin{figure*}
\centering
\includegraphics[width=\hsize]{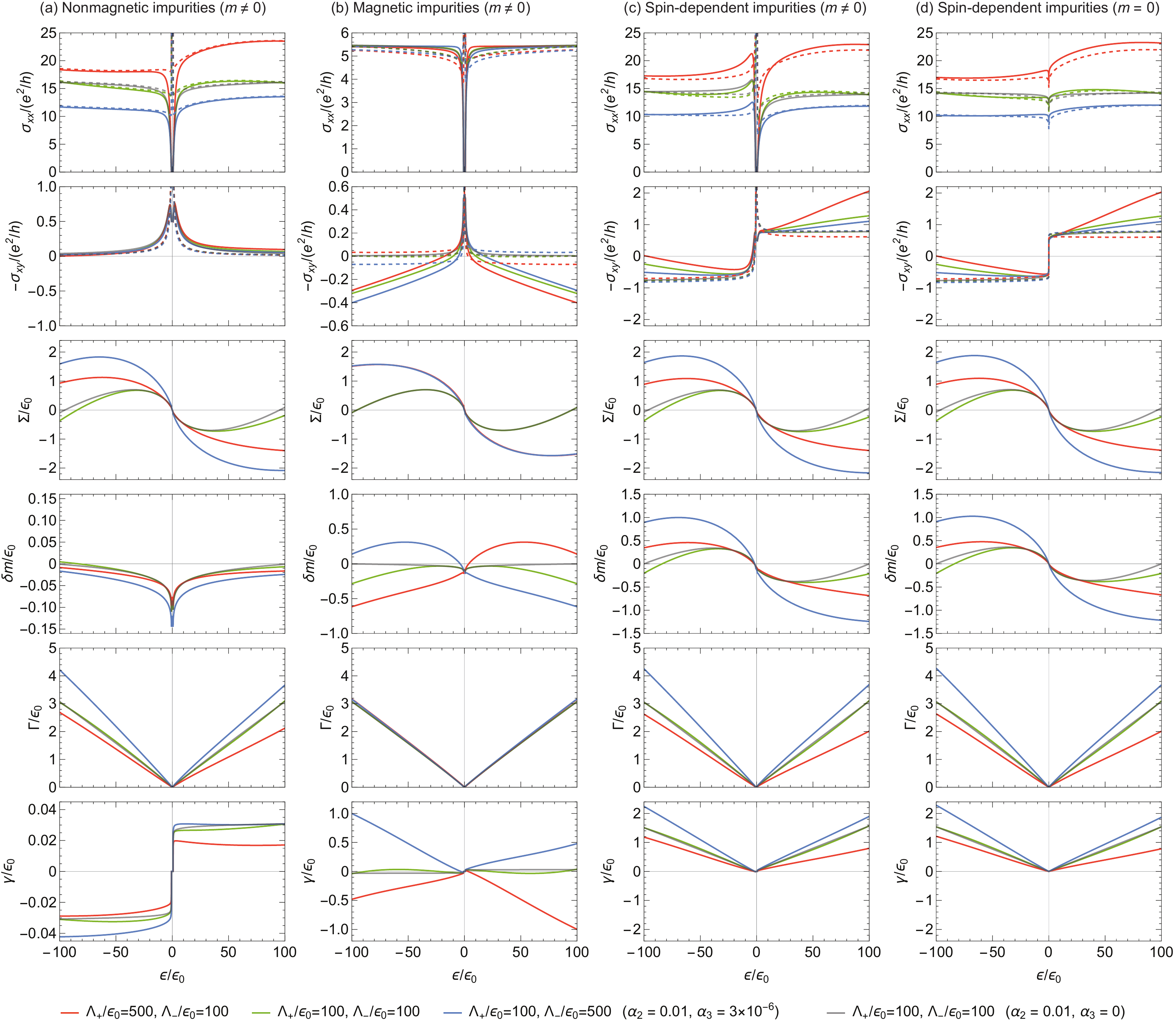}
\caption{
Energy dependence of the conductivity and the self-energy.  
We choose a wider energy range $-100 \leq \epsilon/\epsilon_0 \leq 100$ while using the same parameter set and the legend as those for Fig.~\ref{fig:conductivity}: (a) $\eta = \sigma_0$ (nonmagnetic), $m = \epsilon_0$ (massive); (b) $\eta = \sigma_z$ (magnetic), $m = \epsilon_0$ (massive); (c) $\eta_{11} = \sqrt{3/2}$, $\eta_{22} = \sqrt{1/2}$ (spin-dependent), $m = \epsilon_0$ (massive); and (d) $\eta_{11} = \sqrt{3/2}$, $\eta_{22} = \sqrt{1/2}$ (spin-dependent), $m = 0$ (massless).  
The dimensionless constants for the impurity strength are $\alpha_2 = 0.01$, $\alpha_3 = 3 \times 10^{-6}$ for the colored lines, and $\alpha_2 = 0.01$, $\alpha_3 = 0$ for the gray lines.  Different colors represent different energy cutoffs; see the legend.  
The solid lines represent the exact solutions and the dashed lines the approximate results.  
}
\label{fig:extended}
\end{figure*}

We present the the energy dependence of the conductivity $\sigma_{ab}$ in Fig.~\ref{fig:conductivity}, showing the full results without approximations Eqs.~\eqref{eq:S_sigmaIa}, \eqref{eq:S_sigmaIb}, \eqref{eq:S_sigmaII} and the approximations Eqs.~\eqref{eq:S_sigmaxx_NH}, \eqref{eq:S_sigmaxy_NH} for the energy range $-10 \leq \epsilon/\epsilon_0 \leq 10$.  We show the energy dependence for the wider energy range $-100 \leq \epsilon/\epsilon_0 \leq 100$ with the same parameter set in Fig.~\ref{fig:extended} along with the self-energy.  We note that the graphs include results close to the energy cutoff, where the model may not be justified.  We observe some cases where the approximation $\sigma_{xy}$ is not very good even at large energies because the approximate expression does not fully take account of the effect of skew scattering and higher-order corrections.

\subsection{Self-consistent analytic result at the charge neutrality}

It is hard to find a fully analytic solution, including the self-energy, at an arbitrary energy.  However, we can at least obtain an exact solution using the self-energy derived in Sec.~\ref{sec:S_self-consistent-self-energy}.  It leads to the exact solution at $\epsilon = 0$ with $m = 0$, $\alpha_3 = 0$, and the symmetric cutoffs $\Lambda_+ = \Lambda_-$.  
To obtain the result, we need to examine the limit $\epsilon \to 0$ and find 
\begin{gather}
\bm{\Lambda}(0) = \lim_{\epsilon \to 0} \bm{\Lambda}(\epsilon)
= \bm{1}, \\
\bm{\Gamma}(0) = [ \bm{1} - \alpha_2 \eta_{11} \eta_{22} \bm{\Lambda}(0) ]^{-1} 
= \frac{1}{1-\alpha_2 \eta_{11} \eta_{22}} \bm{1}. 
\end{gather}
Then, the three terms of the conductivity Eqs.~\eqref{eq:S_sigmaIa}, \eqref{eq:S_sigmaIb}, \eqref{eq:S_sigmaII} become
\begin{gather}
\sigma^\text{(Ia)}_{ab} = \frac{e^2}{4\pi^2} \frac{1}{1 - \alpha_2 \eta_{11} \eta_{22}} \delta_{ab}, \\
\sigma^\text{(Ib)}_{ab} = \frac{e^2}{4\pi^2} \delta_{ab}, \\
\sigma^\text{(II)}_{ab} = 0, 
\end{gather}
and the exact expression of the conductivity is 
\begin{equation}
\sigma_{ab} = \frac{e^2}{4\pi^2} \frac{2 - \alpha_2 \eta_{11} \eta_{22}}{1 - \alpha_2 \eta_{11} \eta_{22}} \delta_{ab}.  
\end{equation}
We can use the result to confirm the value at the keen dip of the longitudinal conductivity $\sigma_{xx}(0)$ observed in Figs.~\ref{fig:conductivity}(d) and \ref{fig:extended}(d).

\subsection{Perturbative result}

In the subsection, we confirm our results by comparing them with the previous study by Sinitsyn \textit{et al.} \cite{S_Sinitsyn}, which focused on a limited situation for spin-unpolarized impurities with $\eta = \sigma_0$.  Here, we restrict ourselves to the case for $\epsilon^2 - m^2 > 0$ just to avoid repeatedly appearing $\Theta(\epsilon^2-m^2)$ and consider the impurity scattering perturbatively by assuming $\alpha_2 \gg |\alpha_3 \epsilon / \epsilon_0|, |\alpha_3 (\Lambda_+ - \Lambda_-) / \epsilon_0|$.  Here, we use $\eta_0$ and $\eta_z$ in Eq.~\eqref{eq:S_eta} instead of $\eta_{11}$ and $\eta_{22}$ as independent parameters for a clear presentation:
\begin{equation}
\eta_0 = \frac{\eta_{11}+\eta_{22}}{2}, \quad 
\eta_z = \frac{\eta_{11}-\eta_{22}}{2}.  
\end{equation}

First, we approximate the self-energy.  Under the present approximation, only the imaginary part of the self-energy is important and we can neglect the contributions at order $\alpha_3$. From Eqs.~\eqref{eq:S_Gamma}, \eqref{eq:S_gamma}, we find 
\begin{gather}
\Gamma(\epsilon) \approx \alpha_2 \pi |\epsilon| \left[ (\eta_0^2+\eta_z^2) + 2\eta_0\eta_z \frac{m}{\epsilon} \right], \\
\gamma(\epsilon) \approx \alpha_2 \pi |\epsilon| \left[ 2\eta_0\eta_z + (\eta_0^2+\eta_z^2) \frac{m}{\epsilon} \right].  
\end{gather}
Then, the function $\zeta(\epsilon)$ becomes 
\begin{equation}
\zeta(\epsilon) \approx m^2 - \epsilon^2 - 2i \pi \alpha_2 \left[ (\eta_0^2+\eta_z^2) (\epsilon^2 + m^2) + 2\eta_0\eta_z \epsilon m \right] \operatorname{sgn}(\epsilon).  
\end{equation}
Considering $\operatorname{Im}\ln \zeta(\epsilon) = \operatorname{arg}\zeta(\epsilon) \approx -\pi \operatorname{sgn}(\epsilon)$, we can approximate the ladder function Eq.~\eqref{eq:S_functionLambda} as 
\begin{align}
\bm{\Lambda}(\epsilon) &\approx \frac{1}{(\eta_0^2+\eta_z^2)(\epsilon^2 + m^2) + 2\eta_0\eta_z \epsilon m} \left\{\! \frac{1}{2\alpha_2} (\epsilon^2 - m^2) \bm{1} - 2\pi \!\left[ (\eta_0^2+\eta_z^2) |\epsilon|m + \eta_0\eta_z (\epsilon^2 + m^2) \operatorname{sgn}(\epsilon) \right]\! \bm{\varepsilon} \!\right\}.  
\end{align}

For the vertex function, we should examine the terms with $\alpha_3$ more carefully since the vertex function contains $\alpha_3/\alpha_2$, which could be comparable to $\alpha_2$.  By using the approximate forms 
\begin{gather}
L^\text{R}(\epsilon) \approx \ln\frac{\Lambda_+ \Lambda_-}{\epsilon^2 - m^2} + i\pi \operatorname{sgn}(\epsilon), \\
I^\text{R}_0(\epsilon) \approx -\frac{\Lambda_+ - \Lambda_-}{\epsilon_0} - \frac{\epsilon}{\epsilon_0} \ln\frac{\Lambda_+ \Lambda_-}{\epsilon^2 - m^2} - i\pi \frac{|\epsilon|}{\epsilon_0}, \\
I^\text{R}_z(\epsilon) \approx - \frac{m}{\epsilon_0} \ln\frac{\Lambda_+ \Lambda_-}{\epsilon^2 - m^2} - i\pi \frac{m}{\epsilon_0} \operatorname{sgn}(\epsilon), 
\end{gather}
we find the vertex function Eq.~\eqref{eq:S_functionGamma} to be 
\begin{align}
\bm{\Gamma}(\epsilon) &\approx 
\frac{2[(\eta_0^2+\eta_z^2)(\epsilon^2 + m^2) + 2\eta_0\eta_z \epsilon m]}{(\eta_0^2+3\eta_z^2)\epsilon^2 + (3\eta_0^2+\eta_z^2)m^2 + 4\eta_0\eta_z \epsilon m} \bm{1} \nonumber\\
&\quad - 8\pi \alpha_2 \frac{(\eta_0^2-\eta_z^2) [(\eta_0^2+\eta_z^2)(\epsilon^2 + m^2) + 2\eta_0\eta_z \epsilon m] [(\eta_0^2+\eta_z^2) \epsilon m + \eta_0\eta_z (\epsilon^2 + m^2)]}{[(\eta_0^2+3\eta_z^2)\epsilon^2 + (3\eta_0^2+\eta_z^2)m^2 + 4\eta_0\eta_z \epsilon m]^2} \operatorname{sgn}(\epsilon) \bm{\varepsilon} \nonumber\\
&\quad - 4\pi \frac{\alpha_3}{\alpha_2} \frac{(\eta_0^2-\eta_z^2) (\epsilon^2 - m^2) [(\eta_0^2+\eta_z^2)(\epsilon^2 + m^2) + 2\eta_0\eta_z \epsilon m]}{[(\eta_0^2+3\eta_z^2)\epsilon^2 + (3\eta_0^2+\eta_z^2)m^2 + 4\eta_0\eta_z \epsilon m]^2} \frac{\eta_0m + \eta_z\epsilon}{\epsilon_0} \operatorname{sgn}(\epsilon) \bm{\varepsilon}. 
\end{align}

Then, we obtain the perturbative result of the conductivity from Eq.~\eqref{eq:S_sigmaIa} as 
\begin{gather}
\sigma^\text{(Ia)}_{xx}(\epsilon) \approx 
\frac{e^2}{4\pi^2} \frac{\epsilon^2 - m^2}{(\eta_0^2+3\eta_z^2)\epsilon^2 + (3\eta_0^2+\eta_z^2)m^2 + 4\eta_0\eta_z \epsilon m} \alpha_2^{-1}, \\
\begin{aligned}[b]
\sigma^\text{(Ia)}_{xy}(\epsilon) &\approx 
-\frac{e^2}{\pi} \bigg[ \frac{(\eta_0^2+\eta_z^2) |\epsilon|m + \eta_0\eta_z (\epsilon^2 + m^2) \operatorname{sgn}(\epsilon)}{(\eta_0^2+3\eta_z^2)\epsilon^2 + (3\eta_0^2+\eta_z^2)m^2 + 4\eta_0\eta_z \epsilon m} \\
&\qquad\quad + \frac{(\eta_0^2-\eta_z^2) (\epsilon^2 - m^2) [(\eta_0^2+\eta_z^2) |\epsilon| m + \eta_0\eta_z (\epsilon^2 + m^2) \operatorname{sgn}(\epsilon)]}{[(\eta_0^2+3\eta_z^2)\epsilon^2 + (3\eta_0^2+\eta_z^2)m^2 + 4\eta_0\eta_z \epsilon m]^2} \bigg] \\
&\quad - \frac{e^2}{2\pi} \frac{\alpha_3}{\alpha_2^2} \frac{(\eta_0^2-\eta_z^2) (\epsilon^2 - m^2)^2}{[(\eta_0^2+3\eta_z^2)\epsilon^2 + (3\eta_0^2+\eta_z^2)m^2 + 4\eta_0\eta_z \epsilon m]^2} \frac{\eta_0 m + \eta_z \epsilon}{\epsilon_0} \operatorname{sgn}(\epsilon). 
\end{aligned}
\end{gather}
The result may be regarded as a generalization of Ref.~\cite{S_Sinitsyn}.  
While the so-called intrinsic contribution should not depend on the spin components of the impurities $\eta$, it is worth pointing out that the terms independent of the impurity concentration, namely without $\alpha_p$, show the dependence on $\eta$.  
We note from Eqs.~\eqref{eq:S_sigmaIb}, \eqref{eq:S_sigmaII}
\begin{equation}
\sigma^\text{(Ib)}_{xx}(\epsilon) = O(\alpha_2^0), \quad 
\sigma^\text{(Ib)}_{xy}(\epsilon) = \sigma^\text{(II)}_{xx}(\epsilon) = \sigma^\text{(II)}_{xy}(\epsilon) = 0,
\end{equation}
for $\epsilon^2 - m^2 > 0$. 

To confirm explicitly if our result matches that in Ref.~\cite{S_Sinitsyn}, we set $\eta_0 = 1$ and $\eta_z = 0$, which results in 
\begin{gather}
\sigma_{xx}(\epsilon) = \frac{e^2}{4\pi^2} \frac{\epsilon^2 - m^2}{\epsilon^2 + 3m^2} \alpha_2^{-1} + O(\alpha_2^0, \alpha_3), \\
\sigma_{xy}(\epsilon)
= -\frac{e^2}{\pi} \frac{m}{|\epsilon|} \left[ \frac{\epsilon^2}{\epsilon^2 + 3m^2} + \frac{\epsilon^2 (\epsilon^2 - m^2)}{(\epsilon^2 + 3m^2)^2} \right]
- \frac{e^2}{2\pi} \frac{\alpha_3}{\alpha_2^2} \frac{m\operatorname{sgn}(\epsilon)}{\epsilon_0} \frac{(\epsilon^2 - m^2)^2}{(\epsilon^2 + 3m^2)^2} + O\left( \alpha_2, \frac{\alpha_3}{\alpha_2}, \alpha_3^2 \right). 
\end{gather}
By substituting $\epsilon^2$ with $(vk)^2+m^2$ and using $V_p$ instead of dimensionless $\alpha_p$, we obtain the Hall conductivity  
\begin{equation}
\sigma_{xy} 
\approx -\frac{e^2}{4\pi} \frac{m}{\sqrt{(vk)^2 + m^2}} \left[ 1 + \frac{4(vk)^2}{(vk)^2 + 4m^2} + \frac{3(vk)^2}{[(vk)^2 + 4m^2]^2} \right] - \frac{e^2}{2\pi} \frac{V_3}{n_i V_2^2} \frac{(vk)^4 m}{[(vk)^2 + 4m^2]^2} \operatorname{sgn}(\epsilon), 
\end{equation}
which corroborates our calculations.

\section{Scaling analysis}

\subsection{Scaling by the energy}

\begin{figure*}
\centering
\includegraphics[width=\hsize]{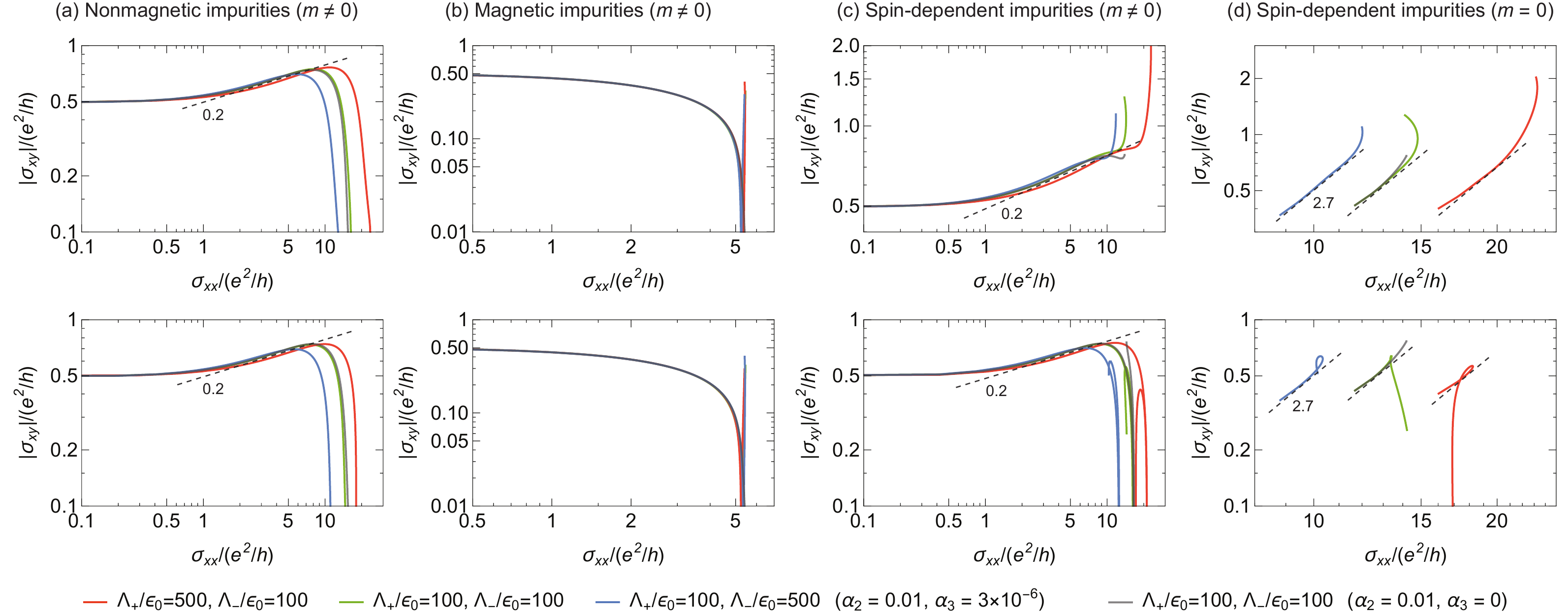}
\caption{
Scaling plots of the conductivity by changing the energy.  We vary energy in the positive range in the upper panels $(0.1 \leq \epsilon/\epsilon_0 \leq 100)$ and in the negative range in the lower panels $(\-100 \leq \epsilon/\epsilon_0 \leq -0.1)$. 
We use the same parameter set and the legend as those for Fig.~\ref{fig:conductivity}: (a) $\eta = \sigma_0$ (nonmagnetic), $m = \epsilon_0$ (massive); (b) $\eta = \sigma_z$ (magnetic), $m = \epsilon_0$ (massive); (c) $\eta_{11} = \sqrt{3/2}$, $\eta_{22} = \sqrt{1/2}$ (spin-dependent), $m = \epsilon_0$ (massive); and (d) $\eta_{11} = \sqrt{3/2}$, $\eta_{22} = \sqrt{1/2}$ (spin-dependent), $m = 0$ (massless).  
The dimensionless constants for the impurity strength are $\alpha_2 = 0.01$, $\alpha_3 = 3 \times 10^{-6}$ for the colored lines, and $\alpha_2 = 0.01$, $\alpha_3 = 0$ for the gray lines.  Different colors represent different energy cutoffs; see the legend.  See Fig.~\ref{fig:extended} for the energy dependence of $\sigma_{xx}$ and $\sigma_{xy}$.  
The dashed lines depict power functions for a guide to the eye with exponents written nearby. 
}
\label{fig:scaling_energy}
\end{figure*}

In Fig.~\ref{fig:scaling_energy}, we show log--log plots of $\sigma_{xx}$ and $\sigma_{xy}$.  We use the same parameter set as that for Fig.~\ref{fig:extended} with the same energy range.  We divide a scaling plot into two for the positive and negative energy ranges to avoid duplicate lines.  
We add some power functions for guides to the eye along with the exponents near the corresponding lines.  There are sharp drops in Figs.~\ref{fig:scaling_energy}(a) and \ref{fig:scaling_energy}(b), reflecting the sign changes of the Hall conductivity $\sigma_{xy}$.  For positive energies in Figs.~\ref{fig:scaling_energy}(c) and \ref{fig:scaling_energy}(d), we observe very rapid increases, of which we cannot determine exponents.  This is partly because the longitudinal conductivity saturates at high energies as we have seen in Fig.~\ref{fig:extended}.  We note that a finite mass becomes negligible at high energies $|\epsilon| \gg |m|$ and that the conductivity behaves similarly there; compare Figs.~\ref{fig:scaling_energy}(c) and \ref{fig:scaling_energy}(d).

\subsection{Scaling by the impurity concentration}

\begin{figure*}
\centering
\includegraphics[width=\hsize]{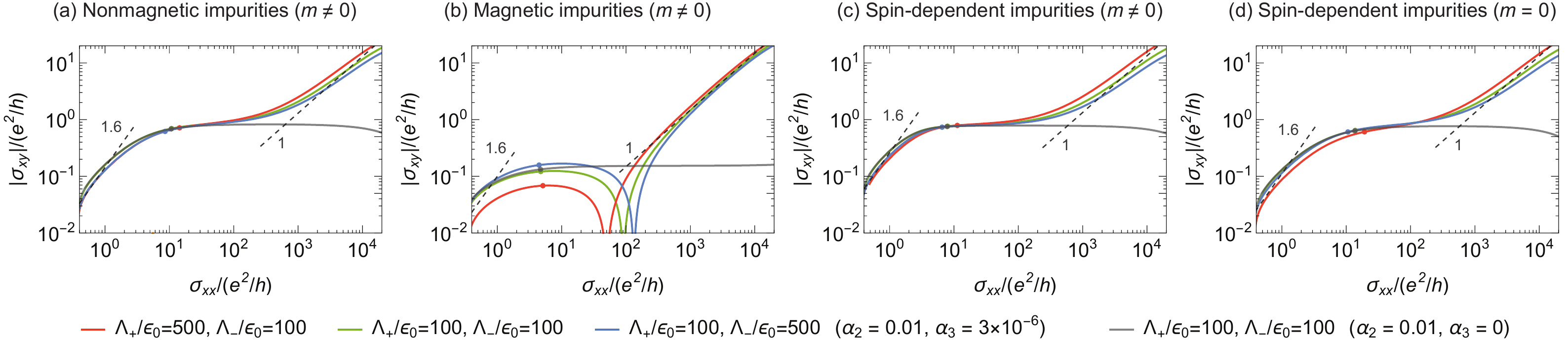}
\caption{
Scaling plots of the conductivity by changing the impurity concentration.  
We set the dimensionless constants for the impurity strength as $\alpha_2 = 0.01 \nu$, $\alpha_3 = 3 \times 10^{-6} \nu$ for the colored lines, and $\alpha_2 = 0.01 \nu$, $\alpha_3 = 0$ for the gray lines.  $\nu = 1$ corresponds to the previous cases as in Fig.~\ref{fig:conductivity}.  The energy is set $\epsilon/\epsilon_0 = 3$.  
We use the same values for the other parameters and the legend as those for Fig.~\ref{fig:conductivity}: (a) $\eta = \sigma_0$ (nonmagnetic), $m = \epsilon_0$ (massive); (b) $\eta = \sigma_z$ (magnetic), $m = \epsilon_0$ (massive); (c) $\eta_{11} = \sqrt{3/2}$, $\eta_{22} = \sqrt{1/2}$ (spin-dependent), $m = \epsilon_0$ (massive); and (d) $\eta_{11} = \sqrt{3/2}$, $\eta_{22} = \sqrt{1/2}$ (spin-dependent), $m = 0$ (massless).  
The dashed lines depict power functions for a guide to the eye with exponents written nearby. 
The points on the lines show the values at $\nu = 1$, corresponding to the results in Fig.~\ref{fig:conductivity} at $\epsilon/\epsilon_0 = 3$.  
}
\label{fig:impurity_scaling}
\end{figure*}

\begin{figure*}
\centering
\includegraphics[width=\hsize]{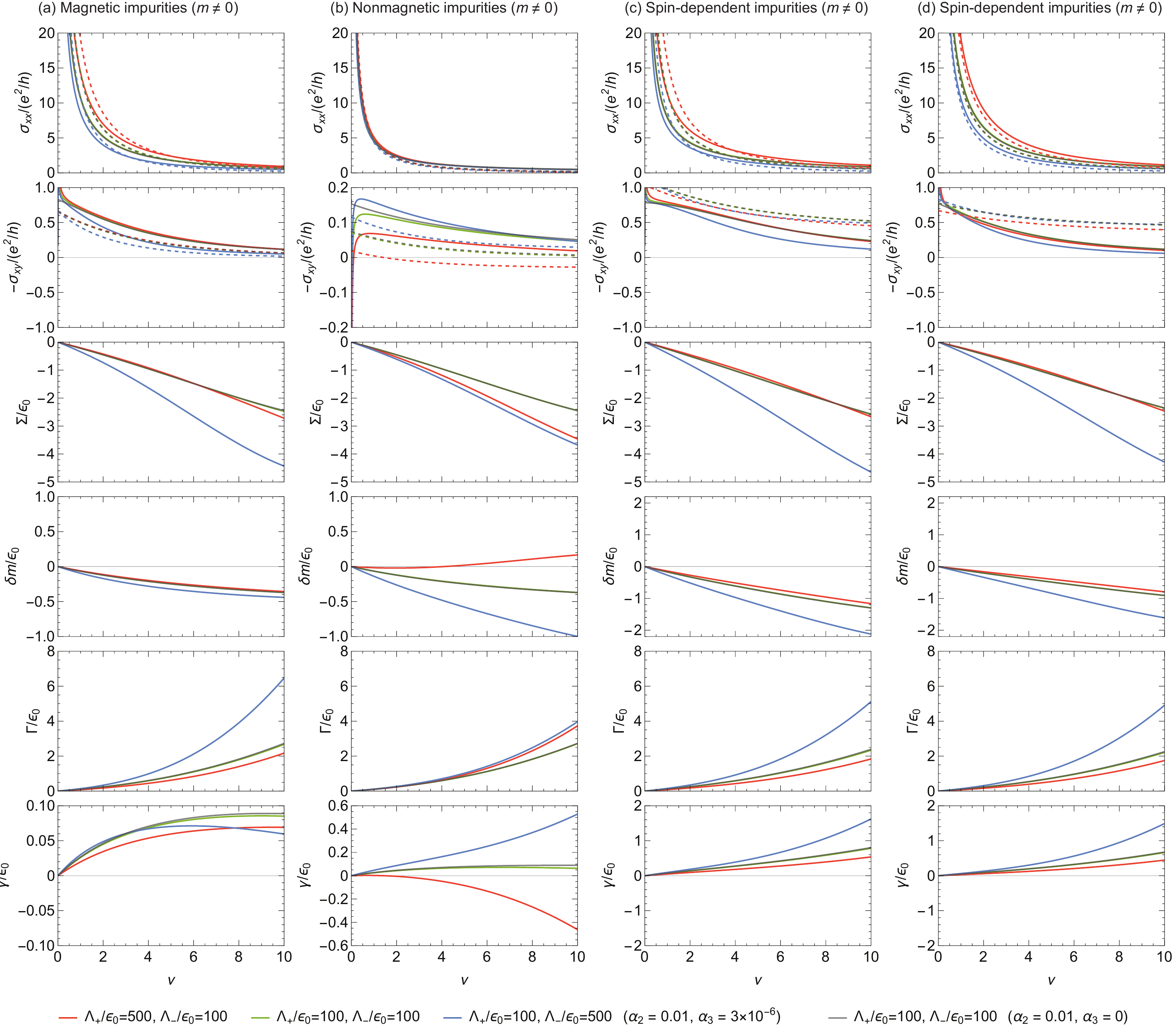}
\caption{
Impurity concentration dependence of the conductivity and the self-energy. 
We parametrize the dimensionless constants for the impurity strength as $\alpha_2 = 0.01 \nu$, $\alpha_3 = 3 \times 10^{-6} \nu$ for the colored lines, and $\alpha_2 = 0.01 \nu$, $\alpha_3 = 0$ for the gray lines.  We vary $\nu$ in the range of $0 \leq \nu \leq 10$ while $\nu = 1$ corresponds to the previous cases as in Fig.~\ref{fig:conductivity}.  The energy is set $\epsilon/\epsilon_0 = 3$.  
We use the same values for the other parameters and the legend as those for Fig.~\ref{fig:conductivity}: (a) $\eta = \sigma_0$ (nonmagnetic), $m = \epsilon_0$ (massive); (b) $\eta = \sigma_z$ (magnetic), $m = \epsilon_0$ (massive); (c) $\eta_{11} = \sqrt{3/2}$, $\eta_{22} = \sqrt{1/2}$ (spin-dependent), $m = \epsilon_0$ (massive); and (d) $\eta_{11} = \sqrt{3/2}$, $\eta_{22} = \sqrt{1/2}$ (spin-dependent), $m = 0$ (massless).  
}
\label{fig:impurity_extended}
\end{figure*}

Instead of varying the energy, we can change the impurity concentration to draw scaling plots.  Whereas the dimensionless constant $\alpha_p$ depends the impurity concentration as $n_i^{p/2}$, we can see that $\alpha_3$ always appears with $\epsilon_0^{-1}$, and hence $\alpha_2$ and $\alpha_3/\epsilon_0$ are both proportional to $n_i$.  We can understand this fact from the original definition of the impurity strength Eq.~\eqref{eq:S_beta}, which arises from the moments of the impurity distribution Eq.~\eqref{eq:S_moments} and is proportional to $n_i$.  
Therefore, changing the impurity concentration from $n_i$ to $\nu n_i$ corresponds to changing the dimensionless constant $\alpha_p$ to $\nu \alpha_p$, while $\epsilon_0$ is kept at the original value.

Figure~\ref{fig:impurity_scaling} depicts the scaling plots by varying the impurity concentration.  We present the four different choices of impurity types and masses, which are the same as those in Fig.~\ref{fig:conductivity}.  We vary the parameter $\nu$ to change the impurity concentration and set the energy $\epsilon/\epsilon_0 = 3$.  We note that the energy unit $\epsilon_0$ does not change while changing the impurity concentration by $\nu$.  
We observe that the scaling plots show the linear dependence for large $\sigma_{xx}$ and the power law with the exponent roughly 1.6 for small $\sigma_{xx}$.  This behavior is similar to the previous study \cite{S_Onoda}, showing skew scattering with the linear power law in the clean limit, side jump with the exponent about 1.6 in a dirty system, and the constant $\sigma_{xy}$ region in the middle.  The origin of constant $\sigma_{xy}$ may be attributed to the intrinsic contribution, which is independent of the impurity concentration.  
We also show the points at $\nu = 1$, where we examined the energy dependence in Fig.~\ref{fig:conductivity}.  Those points lie between the side-jump and intrinsic regions.  
When skew scattering is absent $(\alpha_3 = 0)$, the skew-scattering region does not appear; see the gray lines. 

For a reference, we also present the impurity concentration dependence of the conductivity and the self-energy in Fig.~\ref{fig:impurity_extended}.

\allowdisplaybreaks[0]

\section{Symmetry}
\label{sec:S_symmetry}

\subsection{Symmetries of the mircoscopic model}

We discuss the symmetries of the microscopic model Eqs.~\eqref{eq:S_Hamiltonian} and \eqref{eq:S_impurity}.  
We consider the time-reversal operator 
\begin{equation}
\mathcal{T} = i\sigma_y \mathcal{K}, 
\end{equation}
and the charge conjugation operator 
\begin{equation}
\mathcal{C} = \sigma_x \mathcal{K}, 
\end{equation}
where $\mathcal{K}$ is the complex conjugation operator.  The operators $\mathcal{T}$ and $\mathcal{C}$ are thus antiunitary operators.  
The clean Hamiltonian $H_0(\bm{k})$ has electron-hole symmetry, where $\mathcal{C}$ transforms $H_0(\bm{k})$ as 
\begin{equation}
\mathcal{C} H_0(\bm{k}) \mathcal{C}^{-1} = -H_0(-\bm{k}).  
\end{equation}
On the other hand, finite mass $m$ breaks time-reversal symmetry; the massless system preserves time-reversal symmetry: 
\begin{gather}
\mathcal{T} H_0(\bm{k}) \mathcal{T}^{-1} = H_0(-\bm{k}) \quad \text{with} \quad m=0.  
\end{gather}
We also consider the product
\begin{equation}
\mathcal{S} = \mathcal{T} \mathcal{C} = \sigma_z.  
\end{equation}
It produces another symmetry operation, which we may regard as reflection $(z \mapsto -z)$ perpendicular to the two-dimensional plane.  
We refer to it as ``sublattice'' symmetry for convenience.  
For $m=0$, the clean Hamiltonian satisfies 
\begin{equation}
\label{eq:S_sublattice}
\mathcal{S} H_0(\bm{k}) \mathcal{S}^{-1} = -H_0(\bm{k}) \quad \text{with} \quad m=0. 
\end{equation}
We may regard sublattice symmetry as another (unitary) electron-hole symmetry for $m=0$.  
To summarize, the transformations of the three symmetry operators are 
\begin{alignat}{8}
\mathcal{T}&:&\quad \bm{r} &\mapsto \bm{r}, &\quad \bm{k} &\mapsto -\bm{k}, &\quad \bm{\sigma} &\mapsto -\bm{\sigma}, &\quad \sigma_z &\mapsto -\sigma_z &\quad (z &\mapsto z); \\
\mathcal{C}&:& \bm{r} &\mapsto \bm{r}, & \bm{k} &\mapsto -\bm{k}, & \bm{\sigma} &\mapsto \bm{\sigma}, &\quad \sigma_z &\mapsto -\sigma_z & (z &\mapsto -z); \\
\mathcal{S}&:& \bm{r} &\mapsto \bm{r}, & \bm{k} &\mapsto \bm{k}, & \bm{\sigma} &\mapsto -\bm{\sigma}, & \sigma_z &\mapsto \sigma_z & (z &\mapsto -z). 
\end{alignat}

Symmetries of the system may be different when it has impurities.  If the impurity Hamiltonian transforms similarly as the clean Hamiltonian, the symmetry of the entire system does not change, but if it does not, the symmetry would be lower.  
The three symmetry operators transform the impurity Hamiltonian Eq.~\eqref{eq:S_impurity} as follows: 
\begin{align}
\label{eq:S_impurity_time}
\mathcal{T} V(\bm{r})(\eta_0\sigma_0 + \eta_z\sigma_z) \mathcal{T}^{-1} &= V(\bm{r})(\eta_0\sigma_0 - \eta_z\sigma_z), \\
\label{eq:S_impurity_charge}
\mathcal{C} V(\bm{r})(\eta_0\sigma_0 + \eta_z\sigma_z) \mathcal{C}^{-1} &= -V(\bm{r})(-\eta_0\sigma_0 + \eta_z\sigma_z), \\
\label{eq:S_impurity_product}
\mathcal{S} V(\bm{r})(\eta_0\sigma_0 + \eta_z\sigma_z) \mathcal{S}^{-1} &= +V(\bm{r})(\eta_0\sigma_0 + \eta_z\sigma_z). 
\end{align}
Recall that the boldface symbols signify vectors consisting of $x$ and $y$ components. 

Time-reversal symmetry must be broken for a finite Hall conductivity; therefore either finite mass or magnetic impurities $(\eta_z \neq 0)$ can induce finite $\sigma_{xy}$.  
A massive system only preserves electron-hole symmetry in the clean limit.  We can see from Eq.~\eqref{eq:S_impurity_charge} that magnetic impurities ($\eta_0 = 0$, $\eta_z \neq 0$) also preserves electron-hole symmetry.  
We note that asymmetric energy cutoffs $(\Delta\Lambda \neq 0)$ and the skewness of the impurity potential distribution $(\alpha_3 \neq 0)$ also break electron-hole symmetry.  
Therefore, with magnetic impurities and symmetric energy cutoffs, both longitudinal conductivity $\sigma_{xx}$ and Hall conductivity $\sigma_{xy}$ respect electron-hole symmetry, which we have observed e.g., in Fig.~\ref{fig:conductivity}.

\subsection{Statistical sublattice symmetry}

For a massless case, the clean Hamiltonian preserves time-reversal, electron-hole, and sublattice symmetries, and therefore, magnetic or spin-dependent impurities are necessary to break time-reversal symmetry for a finite Hall conductivity.  We then consider sublattice symmetry.  The operator $\mathcal{S}$ transforms the impurity Hamiltonian as 
\begin{equation}
\mathcal{S} H_\text{imp}(\bm{r}) \mathcal{S}^{-1} = H_\text{imp}(\bm{r}).  
\end{equation}
The impurity Hamiltonian is even under $\mathcal{S}$ while the clean Hamiltonian is odd as we have seen in Eq.~\eqref{eq:S_sublattice}.  
To put it differently, we may regard that the impurity potential $V(\bm{r})$ changes sign under $\mathcal{S}$ while considering it as charge conjugation.  The second-order moment $\alpha_2$ is invariant under $V(\bm{r}) \mapsto -V(\bm{r})$, so that the system \textit{statistically} preserves electron-hole symmetry implemented by $\mathcal{S}$.  We note that this statistical symmetry no longer holds with finite $\alpha_3$.  

As $\mathcal{S}$ virtually flips $z \mapsto -z$, the operation $\mathcal{S}$ forces us to see the system from the opposite side; i.e., from above to bottom.  Then, it changes the sign of the Hall conductivity $\sigma_{xy}$, but not the longitudinal conductivity $\sigma_{xx}$.  
Thus, when the $\mathcal{S}$ symmetry holds statistically, the conductivity satisfies 
\begin{equation}
\mathcal{S}:\ 
\sigma_{xx}(\epsilon) = \sigma_{xx}(-\epsilon), \quad
\sigma_{xy}(\epsilon) = -\sigma_{xy}(-\epsilon).  
\end{equation}
We note that $\eta_z \neq 0$ is necessary for finite $\sigma_{xy}$ when $m=0$.  The result is applicable to Fig.~\ref{fig:conductivity}(d) for $\alpha_3 = 0$.

Electron-hole symmetry by $\mathcal{C}$ also holds with $\eta_0 = 0$.  Since the $\mathcal{C}$ symmetry requires 
\begin{equation}
\mathcal{C}:\ 
\sigma_{xx}(\epsilon) = \sigma_{xx}(-\epsilon), \quad
\sigma_{xy}(\epsilon) = \sigma_{xy}(-\epsilon), 
\end{equation}
$\sigma_{xy}$ vanishes when both $\mathcal{S}$ and $\mathcal{C}$ hold.  Such a case arises when e.g., $m=0$, $\eta_0 = 0$, $\eta_z \neq 0$, $\alpha_2 \neq 0$, $\alpha_3 = 0$.  Then, the Hall conductivity $\sigma_{xy}$ vanishes though time-reversal symmetry is microscopically broken.

\subsection{Time-reversal symmetry for the effective Hamiltonian}

It is evident that the effective Hamiltonian does not respect statistical time-reversal symmetry when the microscopic clean Hamiltonian does not preserve time-reversal symmetry regardless of the details of impurities:
\begin{equation}
\mathcal{T} H_0(\bm{k}) \mathcal{T}^{-1} \neq H_0(-\bm{k}) 
\Rightarrow
\mathcal{T} H^\text{R}_\text{eff}(\bm{k},\epsilon) \mathcal{T}^{-1} \neq H^\text{A}_\text{eff}(-\bm{k},\epsilon).  
\end{equation}
However, even when the effective Hamiltonian preserves statistical time-reversal symmetry, the microscopic Hamiltonian does not always satisfy time-reversal symmetry: 
\begin{equation}
\mathcal{T} H^\text{R}_\text{eff}(\bm{k},\epsilon) \mathcal{T}^{-1} = H^\text{A}_\text{eff}(-\bm{k},\epsilon)
\nRightarrow
\mathcal{T} [ H_0(\bm{k}) + H_\text{imp}(\bm{r}) ] \mathcal{T}^{-1} = H_0(-\bm{k}) + H_\text{imp}(\bm{r}). 
\end{equation}
We can find a microscopic model without time-reversal symmetry but the corresponding effective Hamiltonian satisfies statistical time-reversal symmetry: $m=0$, $\eta_0 = 0$, $\eta_z \neq 0$, $\alpha_2 \neq 0$, $\alpha_3 = 0$.  Then, the previous perturbative results Eqs.~\eqref{eq:S_delta_m} and \eqref{eq:S_gamma} show $\delta m(\epsilon) = \gamma(\epsilon) = 0$ to preserve statistical time-reversal symmetry.  
Interestingly, as we have observed in the previous subsection, the microscopic model is symmetric under $\mathcal{C}$ and $\mathcal{S}$, and therefore, $\sigma_{xy}$ vanishes regardless of time-reversal symmetry breaking.


\begin{thebibliography}{99}
\bibitem{spin}
To relate the Hamiltonian Eq.~\eqref{eq:Hamiltonian} to a surface state of a topological insulator, e.g., Bi$_2$Te$_3$, we may need a spin rotation in the $xy$ plane, which does not alter the discussion.
\bibitem{temperature}
Finite temperature smears the energy dependence and the conductivity becomes $\int d\epsilon [-f'(\epsilon)] \sigma_{ab}(\epsilon)$, where $f(\epsilon)$ is the Fermi--Dirac distribution function.
\end{thebibliography}
\end{document}